\documentclass[12pt,preprint]{aastex}

\newcommand{\etal}{et al.~}
\newcommand{\bperp}{$B_{pos}$}
\newcommand{\intfil}{Integral-shaped Filament}

\shorttitle{Depolarization in OMC-3}
\shortauthors{Matthews, Wilson \& Fiege}

\begin{document}

\title{Magnetic Fields in Star-Forming Molecular Clouds. II. The
Depolarization Effect in the OMC-3 Filament of Orion A} 
\author{Brenda C.~Matthews}
\affil{McMaster University, 1280 Main Street West, Hamilton Ontario, Canada L8S 4M1}
\email{matthews@physics.mcmaster.ca} 

\author{Christine D.~Wilson} 
\affil{McMaster University, 1280 Main Street West, Hamilton Ontario, Canada L8S 4M1}
\email{wilson@physics.mcmcaster.ca}

\author{Jason D.~Fiege}
\affil{Canadian Institute for Theoretical Astophysics, University of
Toronto, Toronto, ON M5S 3H3}
\email{fiege@cita.utoronto.ca}

\begin{abstract}

Polarized 850 \micron\ thermal emission data of the region OMC-3 in
the Orion A molecular cloud are presented.  These data, taken in 1998
with the SCUBA polarimeter mounted on the James Clerk Maxwell
Telescope, have been re-reduced using improved software.  The
polarization pattern is not suggestive of a uniform field structure
local to OMC-3, nor does the orientation of the vectors align with
existing polarimetry maps of the OMC-1 core 20\arcmin\ to the south.
The depolarization toward high intensity regions cannot be explained
by uniform field geometry except in the presence of changing grain
structure, which is most likely to occur in regions of high density or
temperature (i.e.\ the embedded cores).  The depolarization in fact
occurs along the length of the filamentary structure of OMC-3 and is
not limited to the vicinity of the bright cores.  Such a polarization
pattern is predicted by helical field models for filamentary clouds.
Along $\sim 75$\% of the filament's length, the polarization vectors
correlate strongly with the filament axis, a signature of a toroidally
dominated helical magnetic field; however, near the southern cores,
the vectors are offset in direction by $90^\circ$ from the gas
structure of the \intfil, as traced by dust.  We present three
scenarios to explain the observed polarization pattern of OMC-3 in
terms of a helical field geometry.  A helical field which is
toroidally dominated in the north and poloidally dominated in the
south could wrap the filament.  A criss-crossing of two filamentary
structures could produce the observed offset in polarization vectors,
or the filament could be bent into the plane of the sky.  Qualitative
models incorporating a helical field geometry are presented for each
of the latter two cases.

\end{abstract}


\section{Introduction}

One of the outstanding questions in the study of star formation
concerns the relative importance of magnetic fields in the formation
and evolution of clouds, cores and finally protostars.  Magnetic
fields are thought to provide support against gravitational collapse
on large scales, even regulating the filamentary structures observed
within molecular clouds (e.g.\ \citet{fp00a,car97}; Nakamura, Hanawa
\& Nakano 1993).  The process of ambipolar diffusion has been proposed
to regulate the collapse of dense cores to form protostars (see
\citet{shu87}).  However, some magnetic field must be retained within
the protostellar system, since models predict that protostellar
outflows are collimated by magnetic fields
(e.g.~\citet{pud85,uch85,shu94,fie96}).  Girart, Crutcher \& Rao
(1999) present polarization observations of the CO $J=2-1$ line from
outflow of the NGC 1333 IRAS 4A which predict the same magnetic field
direction as their dust polarimetry at 1.3 mm, at least very close to
the outflow source.  Further from the source, the outflow is not
aligned with the inferred field direction, and this could be due to
interaction between the field and outflow.  Alignment between the
outflow and the field of the outflow has been observed in NGC 2024 FIR
5 (Greaves, Holland \& Ward-Thompson 2001) using polarized spectral
line observations.  The most common method of estimating field
strengths has been the detection of Zeeman splitting of atomic and
molecular lines.  Measurement of Zeeman splitting of the \ion{H}{1} 21
cm line provides direct evidence for magnetic fields on very large
scales in the Galaxy and in the envelopes of molecular cloud complexes
such as Orion A \citep{hei87}.  Similar observations with molecular
species within dense cores has proven challenging, with detections
toward only 15 cloud cores \citep{cru99a}.  The sensitivity does not
yet exist to measure the Zeeman effect in more tenous regions of
molecular clouds, where the current generation of polarizers are
beginning to probe.

Emission from aligned, spinning dust grains is anisotropic and hence
polarized.  Polarization data reveal no direct information about the
field strength, since the degree of polarization is dependent on other
factors such as grain shape, composition, and degree of alignment.
The degree of polarization is in essence a measure of how effectively
the grains have been ``sped up'' \citep{hil99}.  Even in theories
where the grain spin is induced by a mechanism other than the magnetic
field, such as the radiation field \citep{dra96} or the production of
H$_2$ on the grain surface \citep{pur79}, the magnetic field is
expected to provide the alignment.  Because of this, continuum
polarization data are the principal means of probing the geometry of
the magnetic field.  Each individual dust grain produces polarized
emission perpendicular to its local field direction.  All polarization
data probe only the plane-of-sky component of the three dimensional
magnetic field (perhaps vexingly denoted $B_\perp$, or \bperp), but
the polarization vectors measured may be either {\it parallel or
perpendicular} to $B_\perp$, depending on whether the polarization
data are due to absorption of background light by dust grains
($\lambda < 25$ \micron), or thermal emission from the grains
themselves ($\lambda > 25$ \micron). \citet{hil88} contains a thorough
review.  At far-infrared and submillimeter wavelengths, dust emission
is optically thin toward all but the densest cores.  Therefore,
observations represent the sum of polarizations contributed by all
dust grains through the depth of the cloud along a line of sight.

Where field geometries are simple and the direction of the magnetic
field does not vary through the cloud depth, the polarized emission
detected is perpendicular to the mean magnetic field and the latter
can be inferred simply by rotating the polarization vectors by
$90^\circ$.  If the field has a more complex, non-uniform geometry,
then interpretation becomes more difficult.  In such cases, it is best
to compare directly the polarization maps with polarization patterns
predicted from a physical model of a magnetized cloud.  For example,
the \intfil\ of Orion A is clearly filamentary, so core models are
inappropriate.  \citet{fp00a} present a model for a filamentary cloud
in which a helical magnetic field threads the filament and plays an
important role in determining the radial density structure.  This
model predicts an $r^{-2}$ density profile, which has been observed in
several clouds, including the \intfil\ \citep{joh99} and several
clouds in Cygnus (Lada, Alves \& Lada 1999; Alves, Lada \& Lada
1999). \citet{fp00c} present predicted polarization patterns for cases
in which the field is either poloidally or toroidally dominated.

Polarimeters aboard the Kuiper Airborne Observatory (KAO), at the
Caltech Submillimeter Observatory (CSO) and the James Clerk Maxwell
Telescope (JCMT) have detected far-infrared and submillimeter
polarization toward many Galactic molecular clouds
\citep{sch97,dow98,sch98,ait00,cop00,dot00,mat00}, although until
recently, limitations on detector sensitivity restricted observations
to bright and/or compact, usually massive, cores.  Detections of
polarized thermal emission from dust have now been made toward
individual protostellar envelopes \citep{gir99,hol96} and starless cores
\citep{war00}.  \citet{dot00} contains a summary of all regions toward
which polarized emission at 100 $\mu$m was detected with the KAO.
Among these sources is OMC-1, a 2000 $M_\odot$ core embedded in the
Orion A \intfil.  These data, along with 350 $\mu$m data from the
CSO's Hertz polarimeter, are presented by \citet{sch98}.  The
polarization pattern observed in OMC-1 has been interpreted as
evidence for an hourglass magnetic field geometry, expected if the
field is being dragged inward with the gas as the core collapses.

\citet{mat00} (hereafter Paper I) presented submillimeter emission
polarization data of OMC-3, a 6\arcmin\ portion of the \intfil\ of
Orion A, located approximately 20\arcmin\ north of OMC-1.  In this
follow-up to Paper I, we present an improved polarimetry map as well
as a broader discussion of the depolarization observed across the
filament.  Additionally, we present three possible explanations for
the observed polarization pattern, two of which require extensions of
the \citet{fp00c} models.  Each of these can potentially explain the
$\sim 90^\circ$ offset from the filament axis observed for
polarization vectors near OMC-3's southern boundary in terms of
magnetic field geometry.  The observations and data reduction
techniques are described in $\S$\ref{obs}.  The polarization data are
analyzed in $\S$\ref{poldata}.  The polarization pattern is
interpreted in $\S$ \ref{interp}, and $\S$ \ref{disc} contains a
discussion and summary.

\section{Observations and Data Reduction}
\label{obs}

Using the SCUPOL polarimeter on the SCUBA detector, 850 \micron\ maps
of polarized thermal emission from dust were obtained on 5 to 7
September 1998 at the James Clerk Maxwell Telescope\footnote{The JCMT
is operated by the Joint Astronomy Centre on behalf of the Particle
Physics and Astronomy Research Council of the UK, the Netherlands
Organization for Scientific Research, and the National Research
Council of Canada.}.  The data set, polarimeter, and general reduction
techniques are described in Paper I, \citet{gre00}, and
\citet{gre01}.  We have re-reduced the data appearing in Paper I using
the Starlink software package POLPACK, designed specifically for
polarization data obtained with bolometric arrays.  The method of
reduction remains consistent with our previous analysis, but the new
software permits easier binning and filtering of data to extract a
higher quality map.  Additionally, noisy bolometers have been removed
from the data and estimates of the instrumental polarization (IP) have
been updated.

At 850 \micron, the sky is highly variable on timescales of seconds.
This variability must be measured and removed from the data.  Chopping
removes the effects of slow sky variability; however, fast variations
remain in the data, requiring sky subtraction using array bolometers
devoid of significant flux.  Typically, we have used between one and
four bolometers (less than the number used in the reduction of Paper
I), depending on the filling factor of the emission across the
detector array.  It is very difficult to select sky bolometers based
on jiggle map data, since bolometers which appear devoid of flux may
in fact have negative fluxes if the chop position contained
significant emission.  This is highly probable in regions of extended
emission such as OMC-3.  To aid in the identification of bolometers
for sky subtraction, we referred to the \citet{joh99} scan maps to
compare our source and chop position fluxes.  If the difference
between these two was approximately zero, the bolometers at those
locations were candidates for use in sky subtraction.  We note that
this method will not be generally available in other regions where
pre-existing scan maps may not exist.  This method prompted us to
exclude from sky subtraction some bolometers used in Paper I.

The methods of sky subtraction are discussed in detail in
\citet{jen98}.  Prior to sky subtraction, images were made to examine
the flux in each bolometer, since bolometers used for sky subtraction
should not have negative values (produced if one has chopped onto a
location with significant flux, for example).  If three bolometers are
used, then the signal per second in those three bolometers is averaged
and subtracted from each bolometer in the map.  Clearly, one should be
left with zero flux on average in the sky bolometers.  If there is
evidence that these bolometers were not completely empty, an estimate
of the total flux removed from the map (by summing all the 1s
removals) can be made and a fraction of that flux is added back into
each bolometer such that the total flux is distributed equally over
all 37 bolometers.  This assures that the total flux in the map before
sky subtraction is the same as after sky subtraction.  In the
re-reduction for this paper, the mean sky level was added back into
the data for OMC-3, thereby avoiding a systematic overestimation of
polarization percentage due to underestimates of the total intensity,
$I$.  For example, in Paper I, a bolometer which contains a flux
almost one quarter that of the MMS8 peak was used for sky subtraction.
In the extreme case where $Q$ and $U$ are unaffected by the sky
removal, this implies that the polarization percentage would be
overestimated by a factor of 1.3.

When dealing with extended sources, such as those in OMC-3, the
observing technique of chopping from source to a reference position
can produce systematic error, since the reference position may not be
devoid of flux and may be polarized.  Appendix \ref{appendixA}
discusses the possible impact such a reference, or chop, position
could have on the measured polarization vectors.  We particularly want
to investigate the so-called ``depolarization effect'' which refers to
the trend, observed in many regions, of measuring lower polarization
percentages at positions of high total intensity emission.  However,
we find that, rather than underestimating the ``true'' source
polarization percentage at high intensity, in fact, chopping onto a
region of polarized emission can produce a systematic increase in
polarization percentage for regions of low intensity.  The degree of
increase varies depending on the ratios between the polarized and
total fluxes at the reference and source positions.  Such an effect
would, of course, appear qualitatively identical to the observed
``polarization holes''.  In the scenario we describe, the
polarizations at high intensity are most representative of the correct
polarization percentage in the source.  Similarly, position angles can
be adversely affected if the polarized emission from the source and
reference positions are oriented differently.  Thus, where chopping
introduces systematic error, the most reliable polarization
percentages and position angles will be observed toward the highest
flux positions.  For the OMC-3 data set, assuming the reference
position is not significantly more polarized than the source, the
analysis of Appendix A suggests that even in the faintest regions of
the OMC-3 map, the maximum error in the position angles introduced by
chopping should range from $10-20$\%.  The slopes of log-log plots of
polarization percentage versus intensity observed should not be
steeper than $-0.3$.

As discussed above, the removal of sky noise is a
critical part of reducing SCUBA polarimetry data.  Appendix
\ref{appendixB} contains a comparison of reductions with and without
sky subtraction for different observations of the same region.
Discrepant results are only obtained if no sky subtraction is
performed.  In fact, whereas non-sky subtracted data tend to show high
uniformity in the polarization vectors across the image, the mean
polarization position angle differs greatly from one observation to
another toward the same region.  Once sky subtraction is applied to
each map, all the resultant polarization patterns show similar trends.

It is also instructive to subdivide the entire data set and compare
one section to another to check for consistency within the data set.
For OMC-3, there are a total of 69 individual polarization
observations, basically centered on four different positions (see
Table \ref{observing_parameters}).  We divided the data evenly, so
that the S/N per area in the map is maintained across each subset (of
34 and 35 observations respectively).  The sky conditions were
extremely stable over all three nights, meaning differences in the
optical depth should have minimal effects on the S/N of the maps.
Polarization maps were then generated for each subset separately,
binned to 12\arcsec\ and compared.  Of the data vectors which had
absolute uncertainties in polarization percentage, $dp$, less than
1.4\% and signal-to-noise in polarization percentage, $\sigma_p$,
greater than 4.2 (values consistent with uncertainties in $dp <1$\%
and $\sigma_p >6$ in the total map), there were 190 vectors in common
between the maps.  Appendix \ref{appendixC} contains maps of each
subset of data.  These maps illustrate that the same general
polarization pattern is produced with each subset.  Comparison of the
percentage polarization data for the 190 vectors in common between the
maps reveals that 70\% of them show insignificant differences between
each other (i.e.\ $(p_1 - p_2)/(dp_1 + dp_2) < 3$ where we have
assumed the errors are correlated in estimating the upper limit on the
uncertainty in the difference).  The largest discrepencies are seen in
the lower intensity regions.

The data presented in this work have lower noise than those in Paper
I.  Binning the data to 12\arcsec\ instead of 6\arcsec\ increases the
signal-to-noise across the whole map by a factor of 2, since binning
is executed in both dimensions.  Data are not thresholded by an upper
bound on polarization percentage as was done for Paper I.  Instead,
thresholding is done on total intensity, uncertainty in polarization
percentage, and signal-to-noise of polarization percentage.  Vectors
with low polarization percentage are suspect due to uncertainties in
IP values of $\pm 0.5$\%, as well as potential sidelobe contamination
(see \citet{gre01b}).  For our regions, polarization percentages less
than 0.5\% are not believable due to sidelobe effects.  Thus, to
account for these two effects, we reject all vectors with polarization
percentage less than 1\%.  The number of vectors presented in this
work is less than that of Paper I, due to higher binning, but the data
quality has increased so that we present polarization data toward
regions of low total intensity, particularly between the cores MMS6
and MMS7.  The mean polarization percentage in Paper I was 4.2\%; the
data presented here have a mean of 5.0\%.  This increase is due to the
polarization vectors present in regions originally used for sky
subtraction.  Their high polarization percentages (particularly in the
southern region) inflates the new average.  We note that even the
bolometers used for sky subtraction in this analysis could also
contain polarized flux; however, we have utilized Appendix
\ref{appendixA} to estimate the potential effects on our data to be
minimal under certain assumptions about the relative polarizing power
off and on the bright peaks.

The most substantial change in the polarization pattern from Paper I
is in the southern region of the map.  As Figure \ref{shift} shows,
the vectors are on average $-30^\circ$ shifted compared with Paper I.
This shift is not exhibited in other regions of the map, and we
performed several tests to eliminate potential sources of the shift.
Using the same bolometers as in Paper I, we still produce distribution
B of Figure \ref{shift}.  We have also tested the effects of the
addition of the mean sky flux level back into the map after the
removal of sky noise, and the distribution remains unchanged.  It is
re-assuring that the distribution of vectors we present in this paper
is robust against these changes in the reduction procedure and that
sky subtraction in POLPACK is not producing the shift.  The remaining
substantive difference between the reductions we have run for this
work and those of Paper I is the use of the POLPACK software itself.
Since Paper I was published before POLPACK was available, the Paper I
solution was the result of a ``brute force'' reduction, which used
straight subtraction of fluxes at different waveplate positions to
generate the $Q$ and $U$ Stokes' parameters.  This method required
substantial thresholding to remove data of poor quality and extract
the polarization percentages and position angles.  Generation of $Q$
and $U$ by subtraction meant that the expected sinusoidal dependence
of the flux with waveplate position was not required.  In POLPACK, the
flux variations with waveplate positions are fit with a sinusoidal
distribution and $Q$ and $U$ are extracted from the fit.  This method
also permits an internal estimation of errors by comparison of
equivalent waveplate positions (i.e.\ $0, 90, 180$ and $270^\circ$
waveplate positions all measure a position angle of $0^\circ$ on the
sky).

The southern region of OMC-3 is distinquished from the rest of the
filament by the presence of extended emission of significant
unpolarized intensity on either side of the filament.  If the
solutions for $Q$ and $U$ were under or over-estimated due to extreme
values in some bolometers, and these were then used for sky
subtraction, then this would explain the systematic shift in the
vectors when comparing the results of Paper I and this work.  It is
not the specific bolometer used which affected the data, but how the
Stokes' parameters were generated which led to the systematic shift.
When POLPACK was used to generate $Q$ and $U$, they were determined
from a best fit to the predicted sinusoidal pattern, thus providing a
much more sophisticated and robust means of extracting the
polarization data. We are confident that the angles presented in this
work are a correct representation of the polarization features in this
region.

\section{850 \micron\ Polarization Data}
\label{poldata}

\subsection{A Non-Uniform Field across OMC-3}
\label{presentdata}

In Paper I, we presented maps of the distribution of position angle
across four regions of OMC-3's filament.  Based on Gaussian fits to
those distributions, we showed that the position angles changed as one
moved down the filament.  In our re-reduced data set, shown in Figure
\ref{mapnew}, data are binned to 12\arcsec\ sampling and bad
bolometers have been removed, thereby increasing the signal-to-noise
in all regions.  The only significant difference is the shift of the
southern vectors to more negative angles as discussed in $\S$
\ref{obs}.

OMC-3 contains ten embedded cores as identified by \citet{chi97}, some
of which show evidence that protostellar collapse has already begun.
However, the polarization pattern is not deflected by the presence of
the dense cores, but is continuous along the filament, as is also
observed in the ridge of cores in the NGC 2024 region of Orion B
(Matthews, Fiege \& Moriarty-Schieven 2001).  This could suggest that
the cores, which have presumably formed by the fragmentation of the
filament, have preserved much of the highly ordered magnetic structure
of the parent filament \citep{fp00b}.  While this may be the case, it
is also likely that, no matter what the field structure in the cores,
the JCMT's 14\arcsec\ beam is insufficient to resolve it, since the
polarization pattern we observe appears to be dominated by the larger
filamentary structure.

It is possible to quantify our argument that the polarization pattern
traces primarily the filament, rather than the dense cores embedded
within it.  The filament's axis was located by fitting a low order
Chebyshev polynomial to the coordinates of the surface flux (or column)
density maxima, equally spaced along the spine.  A low order fit is
desirable since the spine is then defined by a smooth curve which
represents the global structure of the filament rather than responding
in a noisy fashion to each small dense structure traced by dust.  We
then compared the orientation of each polarization vector with the
{\it local} orientation of the filament.  Orthogonal cuts at each
position along the filament were made, and the position angle of the
axis was compared to the position angles of polarization vectors lying
along these radial cuts.  Figure \ref{offsetpa} presents histograms of
the offset angles between polarization vectors and the filament
orientation along its length.  Along most of OMC-3, these
distributions are centred on zero (although the FWHM are large).  We
fit Gaussians to these distributions and found mean offsets of
15$^\circ$, 4$^\circ$ and 1$^\circ$ with $\sigma$ of 26$^\circ$,
21$^\circ$ and 29$^\circ$ for regions MMS1-4, MMS5-6 and the
coreless-MMS7 regions respectively.  South of MMS7, Figure
\ref{mapnew} shows there is a shift in the vector orientation as the
polarization pattern becomes increasingly misaligned with the fitted
north-south spine of the filament.  The distribution of offsets around
MMS8 and MMS9 range from $\sim 60-90^\circ$.  A Gaussian fit to this
distribution yields a mean offset of 86$^\circ$.  So, these vectors
are virtually perpendicular to the filament.

Aside from the basic orientations of the polarization pattern
discussed above, we note here that at the periphery of the detected
polarization data, there are several locations in Figure \ref{mapnew}
where the vectors appear to orient themselves along faint, extended
dust structures.  For example, south of MMS4 lie two small
condensations of dust pointed southward.  Note that the polarization
data lie north-south in this region.  Also, east of MMS6, a faint lane
of dust extends to the north-east, toward the bright source at R.A.\
$05^{\rm h}35^{\rm m}$29\fs9, DEC.\ $-04^\circ$58\arcmin52\farcs7
(J2000).  The region surrounding MMS8 and MMS9 exhibits vector
orientation aligned around $-70^\circ$ (east of north).  This is where
the filament appears widest in OMC-3, with bright peaks at the east
(MMS10, \citet{chi97}) and west boundary of the polarization data.
The suggestion of alignment of polarization vectors with extended
faint dust lanes and the data presented in Figure \ref{offsetpa} for
the region from MMS1 to MMS7 provide strong support for a correlation
between the polarization pattern and dense gas as traced by dust in
that part of the filament.

\subsection{Depolarization Along the Filament Spine}

Diminished polarization percentage toward bright peaks is routinely
observed in extended massive cores, such as those of OMC-1 at 100 and
350 \micron\ \citep{sch98} and even in interferometric maps of the
protostellar source NGC 1333 IRAS4A \citep{gir99}.  In Paper I we
briefly discussed the depolarization observed for a perpendicular cut
across MMS4.  We can now generalize this result for the whole OMC-3
region.  Figure \ref{pvsI} plots the polarization percentage of
vectors with $\sigma_p > 6$ versus the ratio of the intensity to the
peak in the map where the intensities are estimated from the
polarization data.  This figure clearly shows that depolarization
toward higher intensities is a general result in our data set.  The
same behavior is observed in three regions of Orion B \citep{mor01}.

Examination of Figure \ref{pvsI} raises the question of whether the
diminished polarization could be a systematic effect at low values of
total intensity.  The polarization percentage, $p$, is derived from
the Stokes' parameters $Q$ and $U$ where $p = \sqrt{Q^2 + U^2}/I$.
When $I$ is small and $Q\propto I$ and $U\propto I$ are even smaller,
noise effects can lead to gross overestimates of $p$.  Systematic
effects can also be introduced by significant polarization in the chop
position, or reference beam, of the observation, or in the bolometers
selected for sky subtraction (see Appendix \ref{appendixA}).  

Figure \ref{pvsr} shows the depolarization toward the filament axis
for regions containing bright cores as well as the coreless region
between MMS6 and MMS7.  These plots are generated using lines
perpendicular to the fitted slope of the OMC-3 filament discussed in
$\S$ \ref{presentdata}.  The polarization percentage versus (the
magnitude of) the radial distance from the filament axis along these
``radial cuts'' are plotted.  No interpolation is done; each plotted
point is a true data point on Figure \ref{mapnew}.  Most of the cores
(with the exception of MMS4) show increasing polarization at greater
radial distances.  The trend of declining polarization toward the axis
is not limited to the bright cores embedded in the filament.  Figure
\ref{pvsr} shows that depolarization persists along the length of the
filament spine, even in a region devoid of bright cores between MMS6
and MMS7.  The depolarization becomes deeper as one moves southward,
including the coreless region.  The fact that depolarization is
observed along the entire length of OMC-3 is further evidence that the
polarization pattern is a feature of the filament itself and does not
require the presence of dense, cold cores.  Additionally,
depolarization along the filament suggests that any model of
the polarized emission from magnetized filamentary clouds must be able
to explain the presence of depolarization toward the central axis.

Thus, we conclude that the depolarization effect is a feature of the
filament, not the dense cores, although steeper depolarization may be
observed near cores due to augmented effects of field tangling on
scales smaller than the beam or more distinct grain changes.  While
the depolarization effect is a signature of a helical field, we do not
suggest that field geometry is the sole means by which such an effect
could be produced.  Other factors which can produce a depolarization
hole include systematic effects (i.e.\ chopping onto polarized
emission as discussed in Appendix \ref{appendixA}), although we do not
think the flux levels in these maps could reproduce the depth of
depolarization we observe.  One possible explanation could be that the
grain physics is changing with proximity to the axis of the filament,
i.e.\ with density or optical depth \citep{hil99}.  If the degree of
alignment or spin rate changes with density, then the grains near the
central axis could exhibit a lesser degree of polarization.  At higher
column densities, the grains could become more spherical through
agglomeration processes, thereby rendering them unpolarizable.  If
these changes are present at the densities and temperatures
characteristic of the center of the filament, this too could explain,
or at least contribute to, the depolarization effect.

\subsection{Can Faraday Rotation Account for Depolarization?}

Faraday rotation occurs if a linearly polarized wave encounters a
plasma containing a magnetic field (i.e.\ the ISM).  A linearly
polarized wave can be decomposed into two circularly polarized waves
of opposite orientations. These components propagate with different
phase velocities in the plasma, if there is a component of the
magnetic field along the line of sight.  The net effect will be one of
depolarization if the cloud is optically thin and field strength and
electron densities are such that Faraday rotation is significant.
Since thermal dust emission toward all molecular regions (with the
exception of the central regions of dense collapsing cores) is
optically thin, the net polarized emission measured by the telescope
will be a sum over all emitting grains subject to different degrees of
Faraday rotation depending on the different pathlengths through which
they traveled through the cloud.  Thus, vector addition of Faraday
rotated emission from the far side of a cloud and virtually unaffected
emission from the near side of the cloud could produce a net
polarization vector of zero.

The plane of polarization of the linearly polarized wave rotates as it
passes through the plasma by an amount:
\begin{equation}
\Delta \theta = RM \ \lambda^2 \ \ \ \ \ \rm{radians}
\label{change_angle}
\end{equation}

\noindent where $\lambda$ is the wavelength of the observation in
meters and $RM$ is the rotation measure, given by:
\begin{equation}
RM = (8.1 \times 10^5) \int n_e B_{||} dr \ \ \ \ \ \rm{rad \ m^{-2}}
\label{RM}
\end{equation}

\noindent where $B_{||}$ is the magnetic field strength along the line
of sight in Gauss; $n_e$ is the number of elections per cm$^{-3}$; and
$\int dr$ is the path length in parsecs \citep{kra86}.  Although we cannot deduce a
rotation measure from our data, it is useful to demonstrate that
Faraday rotation has a negligible effect on the orientation of
polarization vectors presented in Figure \ref{mapnew}.  The analysis
can be broken into two parts: the possible effects of Faraday rotation
in the ambient, diffuse ISM toward Orion; and within Orion A itself.

In the former case, we can derive $\Delta \theta$ at 850 $\mu$m by
using the median value deduced from 81 extra-galactic sources at the
high-longitude boundaries of the Canadian Galactic Plane Survey at 21
cm.  For the vast majority of these sources, $RM$ values between
$-400$ and 50 rad m$^{-2}$ have been measured, with a median value of
$-163$ rad m$^{-2}$ \citep{bt01}.  For $\lambda = 850 \ \mu$m, this
$RM$ implies an angular change of $-1.2 \times 10^{-4}$ rad or
$-0.007^\circ$ of rotation along the whole line of sight through the
Galaxy.  Thus, any rotation effects of the ISM between the Sun and
Orion, a mere 500 pc away, are negligible.

Within the Orion A cloud, an estimate of Faraday rotation is more
difficult to deduce.  Measurements have been made of the field
strength toward the dense ($10^6$ cm$^{-3}$) CN core of OMC-1 of
$B_{||} = -0.36 \pm 0.08$ mG \citep{cru99}.  Since the CN measurements
sample gas up to 100 times denser than that sampled on large scales
along OMC-3 at 850 \micron, it is unlikely that the field strengths
local to the gas sampled by our data would exceed 35 $\mu$G at $10^4$
cm$^{-3}$.  Assuming an ionization fraction of $10^{-4}$ (i.e.\
\citet{ung97}), implies that $n_e \approx 1$ cm$^{-3}$ local to the
gas sampled by the JCMT.  At its widest point, the filament in OMC-3
is approximately 225\arcsec\ which corresponds to 0.5 pc.  If its
depth is of comparable scale, and if $B_{||}$ and $n_e$ are assumed to
be constant, equation (\ref{RM}) yields $RM = 14$ rad m$^{-2}$ within
the OMC-3 filament.  Since the OMC-3 filament is clearly denser than
the ambient molecular region around it, we assume it would be the
strongest source of Faraday rotation, having higher $n_e$ and $B_{||}$
values than its surroundings.  Substitution in equation
(\ref{change_angle}) yields $\Delta \theta = 6 \times 10^{-4}$ degrees
of rotation for the polarimetric angle through OMC-3.  In regions
denser than $10^4$ cm$^{-3}$, rotation would be proportionally higher,
although the depth ($dr$) of such regions would become progressively
smaller.  Hence we conclude that the effects of Faraday rotation
cannot be responsible for the depolarization effect observed in OMC-3.

\section{Interpreting the Polarization Pattern}
\label{interp}

As we concluded in Paper I, the re-reduction of data toward OMC-3
reveals no evidence for the presence of a uniform field on large
scales across the filament since there are two distinct subsets of
data -- one aligned with and one orthogonal to the filament.  Our
analysis shows that along $\sim 75$\% of the OMC-3's projected length,
the polarization pattern follows the orientation of the filament,
becoming misaligned only south of MMS7.  Further analysis reveals that
the depolarization effect toward OMC-3 is a global feature of the
region, existing along the entire length of the spine.  Therefore, any
theoretical models for this filament should support variations in
field geometry and explain the depolarization effect in the absence of
embedded cores.

In the far-infrared and submillimeter regimes, theoretical modelling
of dust grains suggests that each grain along a line of sight
contributes thermal radiation polarized perpendicular to the local
direction of the magnetic field in the plane of the sky \citep{hil88}.
As a result, polarimetry observations have often been interpreted by
rotating the observed electric field vector orientations by 90$^\circ$
to estimate the magnetic field direction.  However, since magnetic
fields are inherently three dimensional, there may exist different
configurations which can produce similar two dimensional polarization
patterns.  Since models of magnetized filamentary clouds are now
available \citep{fp00a,car97,nak93}, a reasonable approach is to vary
magnetic field parameters in a model and then generate the expected
polarization pattern to compare to observations \citep{fp00c}.  Note
that the Fiege \& Pudritz model employs an axisymmetric magnetic
field, so that the field is helical in general.  However, their model
is also consistent with filaments threaded by purely poloidal fields,
although such a geometry is not supported by this data set (see \S
\ref{disc}).

Some basic successes of the Fiege \& Pudritz model include its
prediction of an $r^{-2}$ density profile for filaments, which has
been observed in Orion A \citep{joh99} and two dark clouds in Cygnus
\citep{all99,lal99}.  The model also predicts that the depolarization
observed along the axis of the filament is a natural result of the
helical field geometry and does not rely on poorly polarizing or
poorly aligned grains at high optical depths although the field
geometry certainly does not preclude the existence of such effects.
The basic idea is that polarization contributions from the poloidally
dominated axis of the filament partially cancel the contributions from
the toroidally dominated envelope.  The cancellation is greatest along
the axis, creating the depolarization observed.  However, we note that
poorly polarizing or unaligned grains could act in concert with the
helical field to amplify the depolarization.

The northern region of OMC-3 bears a strong resemblance to the inner
regions of the Type 1 models of \citet{fp00c}, for which
$B_{z,S}/B_{\phi,S}$, the ratio of the poloidal to toroidal field
components at the surface of the filament, $\le 0.1$.  Note that
$B_z/B_\phi$ is at a minimum at the surface of the filament and is
typically $> 1$ in the central regions.  The models presented in
\citet{fp00c} use a maximum polarization percentage of 10\%, on the
order of what we observe in OMC-3 and in three regions of Orion B
\citep{mor01}.  Also, the width of the expected polarization hole
predicted by the \citet{fp00c} model varies with $B_{z,S}/B_{\phi,S}$,
increasing as a function of filament diameter as the relative poloidal
strength increases.  According to these models, the ratio of the width
of the polarization hole to the filament diameter should be 0.5 or
less for Type 1 filaments.  Therefore, more sensitive measurements
with longer chop throws (or no chopping at all) should detect a
decline in polarization percentage at larger radial distances from the
filament, if there is no significantly magnetized medium external to
the filament.  The only region where this effect is suggested by our
data is between MMS6 and MMS7.  Figure \ref{pvsr}c shows
depolarization toward the axis and a single vector of declining
polarization percentage at approximately 35\arcsec\ from the axis.
Figure \ref{mapnew} shows that there are smaller polarizations below
the $\sigma_p = 6$ level, but these are not preferentially further
from the filament than those of Figure \ref{pvsr}c.

Polarization vectors perpendicular to the filament axis are predicted
for poloidally dominated field patterns (see Fig.\ 1 of
\citet{fp00c}).  For filaments symmetric about a central axis, only
vectors parallel or perpendicular to filament axes are expected
\citep{fp00c}.  Paper I reported a misalignment of the polarization
vector position angles of 35$^\circ$ - 47$^\circ$ near MMS8 and MMS9
from the estimated filament orientation of 0$^\circ$ (east of north).
Re-reduction and direct fitting of the filament spine yields a new
estimate of 86$^\circ$ as the difference between the position angles
of the vectors and the orientation of the filament in this region (see
$\S$ \ref{presentdata}).  The fact that this misalignment occurs
within the boundaries of a single SCUBA field of view raises concern
that a systematic effect in observing technique could be producing
these vectors.  In Appendix \ref{appendixA}, it is shown that flux in
the reference position can have detrimental effects on the measured
polarization percentage and position angle.  However, even in extreme
cases of significant polarized flux in the reference position, errors
of 90$^\circ$ in position angle can only be produced for observed
intensities close to zero and are extremely unlikely unless the
polarization percentage in the reference position exceeds that of the
source.  Thus, it is unlikely that this source of systematic error is
responsible for the polarization angles observed.

\citet{fp00c} consider both toroidally and poloidally dominated field
geometries.  In the northern part of OMC-3, the alignment of the
polarization vectors along the filament agrees well with the
predictions for a toroidally dominated field geometry.  However, a
poloidally dominated field is expected to produce a polarization
pattern offset by 90$^\circ$ in position angle from the filament axis.
The poloidally dominated pattern thus predicts the position angles
observed in the southern part of OMC-3.  However, there are several
marked differences between the predicted poloidally-dominated pattern
and the observed vectors.  The predicted pattern has a local maximum
in polarization percentage along the axis, with two symmetric
depolarization holes on either side (c.f.\ Fig.\ 1 of \citet{fp00c}).
The polarization is then seen to rise again at larger radii from the
axis.  However, Figure \ref{pvsr}b does not exhibit this behavior.
The lowest values of polarization percentage measured are along the
axis, just as in the rest of OMC-3.  The cores to the east and west
and the widening of the filament here makes this area difficult to
model.  Further study of this portion of the OMC-3 region with a
larger spatial scale map should provide more insight into the
possibility that this region is poloidally, rather than toroidally,
dominated.

Models invoking a purely poloidal magnetic field geometry aligned with
the axis cannot be reconciled with the polarization pattern along the
northern part of OMC-3.  Additionally, an $r^{-4}$ profile is
predicted by the classic unmagnetized, isothermal filament of
\citet{ost64}.  In fact, it can be shown that all isothermal models
limited to poloidal field geometries and constant flux-to-mass loading
along the field lines produce density gradients steeper than $r^{-4}$.
In the \intfil, \citet{joh99} measure a profile of $r^{-2}$ as
predicted for a helical field geometry.

When the vectors are overlain on the total intensity maps generated
from the polarization data (the sum of the maps obtained for each
waveplate position), the extent of the maps is limited to the SCUBA
fields observed with the polarimeter (c.f.\ Fig.\ 1 in Paper I).
However, larger scale 850 $\mu$m scan maps of this region
\citep{joh99} allow us to place the polarization data in a broader
context since they can be compared to larger scale dust features of
the region. Close examination of the greyscale intensity of Figure
\ref{mapnew} suggests two possible explanations for the observed
$90^\circ$ offset of the polarization vectors from the filament in the
southern region of OMC-3.  The dust emission becomes very extended and
diffuse around MMS8-9.  The continuum source MMS10 also lies to the east
of the \intfil.  A second (un-named) peak could lie to the west of
MMS8.  These bright sources, coupled with the extended low intensity
emission, suggest that a second filamentary structure, nearly
orthogonal to the main filament, could be present.  In this case, the
polarization vectors are in fact aligned with a filament axis, but not
that of the \intfil.  Figure \ref{crossed} shows a qualitative
illustration of the effect of crossed filaments, both of which are
threaded by helical fields.  The second filament has half the central
density of the main filament, which runs roughly north to south.  The
filaments intersect only in projection in this model; at the projected
overlap, the vectors align with the second (east-west) filament.  The
\citet{fp00c} model defines three free parameters.  The first is a
concentration parameter, $C$, given by $C=log (r/r_0)$ where $r$ is
the radius and $r_0$ is the core radius within which the density
profile is flat.  The core radius is given by $r_0 = \sigma (4\pi G
\rho_c)^{-0.5}$ where $\sigma$ is the one-dimensional line width, $G$ is the
gravitational constant and $\rho_c$ is the central density.  Both
filaments have a concentration parameter, $C$, of 1.2, and
dimensionless flux-to-mass loading parameters of $\Gamma_z =13$ and
$\Gamma_{\phi} = 18$ (as defined in \citet{fp00c}).

If a second filament is present, its effects on the polarization
pattern should obviously be limited to its width.  This means that
more extensive polarization observations south into OMC-2 should show
re-alignment with the \intfil.  The second filament also appears to
extend in total intensity beyond the polarization data of Figure
\ref{mapnew} to the northwest and southeast.  Extended polarization
data in these regions should reveal polarization data well-aligned
with this faint dust emission, if two crossed filaments are present.
Continued alignment with the dust structures will provide further
evidence that the field orientation is related to the dense gaseous
structures, either because the magnetic field has guided the
condensation of gas, or because the gas has dragged in the magnetic
field as the filament formed.

A second possibility is that the \intfil\ is the only filament
present, but that it is bent (and thus changes inclination) south of
MMS7.  As discussed above, the northern region agrees well with the
Type 1 models of \citet{fp00c} for a filament in the plane of the sky.
However, even inclination of such models out of the plane of the sky
is not expected to produce position angles other than 0$^\circ$ or
90$^\circ$ on the sky.  If the filament were bent, however, then the
cylindrical symmetry would be broken, and different position angles
can result.  This effect is easily considered qualitatively in terms
of a wrapping cylindrical shape, such as a slinky, where the slinky
represents magnetic field lines at a particular radius where the field
is toroidally dominated.  Consider what happens to the $B_\phi$ loops
when you bend the slinky: they are compressed on the inside of the
bend and pulled apart on the outside.  Due to concentration of the
magnetic field, the inside part dominates, breaking the front-back
symmetry in the straight filament models and causing the vectors to
turn in the direction orthogonal to the field in the inner part of the
bend.  One can thus consider the effect on an observed polarization
pattern in projection.  As long as there is cylindrical symmetry,
vectors in front of and behind the axis can be paired and will sum to
either 0$^\circ$ or 90$^\circ$.  However, once bends are introduced,
then the components of the vectors in the plane of the sky can be very
different on either side of the filament axis, and the projected
vectors cannot be paired.  In this scenario, vector sums through the
cloud may take on any value.  Figure \ref{bent} shows an example of
such a model, where we self-consistently bend both the filament and
the helical field using a Lagrangian formulation of the induction
equation in the limit of perfect MHD.  A full description of our
technique will be presented in a forthcoming publication.  As for the
crossed filament model discussed above, this model filament has a
concentration parameter, $C$, of 1.2, and dimensionless flux-to-mass
loading parameters of $\Gamma_z =13$ and $\Gamma_{\phi} = 18$ (see
\citet{fp00c}).

\section{Discussion}
\label{disc}

We have re-reduced the \citet{mat00} 850 $\mu$m polarization data of
the OMC-3 filament in Orion A, a region exhibiting strong filamentary
structure and undergoing active star formation.  Polarization
observations are the key to revealing the presence of ordered magnetic
fields in star-forming regions and determining whether their geometry
is correlated with regions of high gas densities, as traced by dust.
In OMC-3, we observe strong alignment between the polarization data
and the orientation of the filamentary dense gas in the north,
regardless of where embedded cores are located.  Near the edges of our
polarization data set, vectors appear to rotate to coincide with the
orientations of faint structures of lower densities as illustrated in
a larger scale intensity map of \citet{joh99}.  Near the southern part
of OMC-3, the vectors rapidly shift orientation, becoming almost
orthogonal to the orientation of the bright \intfil, which could
indicate the presence of a poloidally dominated field there.

Optical absorption polarization data on the periphery of the Lynds
1641 cloud \citep{vrb88} and in M42 \citep{bre76} reveal a net
polarization direction of 120$^\circ$.  In the case of a uniform
field, we would thus have expected emission polarization data to
present vectors oriented at a position angle of $\sim 30^\circ$ (since
absorption and emission polarimetry should be orthogonal if they trace
the same field geometry).  The 100 and 350 $\mu$m data of OMC-1,
located 20\arcmin\ further south along the \intfil\ from OMC-3,
exhibit a polarization pattern with a mean direction of approximately
30$^\circ$ east of north (Schleuning 1998).  These data were interpreted as
support for a uniform field with position angle $120^\circ$ (east of
north) throughout the whole Orion A cloud.  However, none of the
vectors in OMC-3 suggest such a field geometry.  Although the
polarization data in the northern part of OMC-3 vary smoothly, they
are not aligned with the data of OMC-1 nor the large scale optical
data.  In fact, the polarization orientations differ by $70-80^\circ$.
Furthermore, the southern part of OMC-3 shows an abrupt change in
polarization orientation, which is not easily explained by a uniform
field.  Changing vector orientations (i.e.\ south of MMS7) could
indicate a bend in the field lines.  This is why the polarization
pattern from the OMC-1 core was interpreted as being pinched in due to
collapsing gas \citep{sch98}.  The mean position angle near MMS8-9
is not consistent with that of either OMC-1 or northern OMC-3.  If the
magnetic field is uniform (i.e.\ of identical strength and direction
throughout the depth of the cloud), all the vectors should line up in
the same direction, regardless of the behavior of the gas.

Thus, taken as a whole, the data in OMC-3 alone are not consistent
with a uniform field.  Including the OMC-1 core data as well brings
the number of ``mean field directions'' in these two data sets up to
at least three.  Under the picture of a uniform field, this should not
be the case.  Interestingly, the orientation of the filament in OMC-1
can be estimated using the alignment of the two brightest cores.  The
angle between them is $\sim 30^\circ$ east of north, with is
consistent with the polarization position angle measured by
\citet{sch98} on large scales, but not with the interferometric
position angles measured at 1.3 and 3.3 mm with the Berkeley Illinois
Maryland Association (BIMA) interferometer \citep{rao98}.

An indirect method of estimating the magnetic field strength from
polarimetry utilizes the assumption that the dispersion in the
position angles of vectors is related to the magnetic field strength
\citep{cha53}.  Inherent in this method is the assumption that there
is a mean field orientation which can be identified (such as in the
case of the spiral arms of the Galaxy in the original work of
Chandrasekhar and Fermi).  For complex field geometries in which the
field reverses, there is no mean field to define; hence, we do not
utilize this method to estimate a field strength toward OMC-3.

Our basic interpretations of the polarization pattern in Paper I
remain unchanged.  A comparison of filament to polarization position
angles shows distributions centered on zero from MMS1 to MMS7, below
which the vectors slowly rotate until they are misaligned from the
filament (with a mean offset of $86^\circ \pm 19^\circ$).
Additionally, the vector orientations are inconsistent with those
predicted for poloidal fields where vectors would align perpendicular
to the filament \citep{fp00c}.  The northern data could suggest the
presence of a transverse field.  However, this interpretation would
require either poorly aligned or poorly polarizing grains near the
central axis of the filament.  From a dynamical perspective, one might
also expect a flattened sheet rather than a filament for this field
geometry, since the magnetic support would be in a plane orthogonal to
the field direction.  \citet{hei87} observed Zeeman splitting of
\ion{H}{1} in the atomic envelope of Orion A and found evidence which
supports the presence of a helical field geometry, although it was
later interpreted in terms of the expanding Eridanus loop
\citep{hei97}.  The \ion{H}{1} data sample a different gas component
of the ISM than that probed by our polarimetry.  Therefore, in order
to interpret the three-dimensional field geometry local of the
\intfil, it is vital that Zeeman splitting measurements of molecular
gas within the filament be obtained.

In Paper I, we showed only a single radial cut across the MMS4 core to
illustrate the depolarization across one of the filament's bright
cores.  A logarithmic plot of $p$ vs.\ $I$ for our entire data set
shows that depolarization is a {\it global} feature in this filament.
Furthermore, the distribution of the polarization percentage as a
function of the distance from the filament spine reveals that
depolarization exists along the entire length of OMC-3, most
importantly across the coreless region between MMS6 and MMS7.  This
result implies that depolarization is a feature of this filament even
in the absence of condensed cores.  Therefore, any model of
filamentary clouds must be able to explain this feature.

The existing polarization data toward OMC-3 are insufficient in
spatial extent to discriminate between two of our proposed
explanations of a second filamentary structure or a bending in the
\intfil.  The behavior of the polarization south of MMS9 (into the
region of OMC-2) is of particular interest.  If the vectors in OMC-2
behave as the northern pattern of OMC-3, this could indicate support
for a second filament, since only the region of juxtaposition is
affected.  (Observations of this region in CO with high velocity
resolution will help to resolve the question of whether two filaments
are juxtaposed on the sky.)  On the other hand, if the vectors are
asymmetric or misaligned from the parallel or perpendicular
orientations in OMC-2, then this could indicate the model of a single
bent filament better represents the physical properties of the
\intfil.  Extending the map to the east and west will reveal if strong
polarization continues along the faint emission around MMS8 and MMS9.
If neither hypothesis is supported by extending the data set, the
filament may truly be poloidally dominated near the MMS8 and MMS9
cores.

In Paper I, we speculated that the effects of outflow from the
powerful class 0 source MMS9 could have affected the magnetic field
geometry to the east and west of that source, producing the
polarization pattern observed.  This is a particular concern for 850
\micron\ SCUBA polarimetry because the $^{12}$CO $J=3-2$ line lies at
the center of the 850 \micron\ filter bandwidth.  The CO line
may also be polarized and, if the CO emission is significant, it can
dominate the polarization of the continuum.  However, the only part of
OMC-3 where the CO has been shown to dominate the 850 \micron\
continuum is in a Herbig-Haro knot west of the filament \citep{joh01}.

Even if the field of an outflow is aligned with that of the young
stellar object, it does not follow that the polarization directions
would be the same.  For example, \citet{gir99} found orthogonal
polarization directions from 1.3 mm continuum dust in NGC 1333 IRAS 4A
and $^{12}$CO $J=2-1$ line in its outflow.  Finally, the outflow of
MMS9 does not extend as far north as MMS7, where the direction of the
polarization vectors begins to change.  New data on outflows in the
OMC-2/3 region identify MMS9 as the primary driving source in the
southern part of OMC-3, while MMS8 is not associated with outflow in
either $H_2$ shocks or CO emission \citep{aso00,yu00}.  The outflow
from MMS9 is aligned in a northwest to southeast orientation, contrary
to the northeast to southwest orientation of the continuum emission.
Based on CO emission, \citet{yu00} claim that MMS10 as identified by
\citet{chi97} has no submillimeter counterpart.  However, in the 850
\micron\ map of \citet{joh99} shown in Figure 1, there is a peak
coincident with the 1.3 mm dust condensation observed previously.
\citet{aso00} identify MMS10 as driving an east-west outflow.
Disentangling the magnetic signatures of the cores and outflows in
this region will require direct measurement of the polarization of the
$^{12}$CO $J=3-2$ or the $^{12}$CO $J=2-1$ line.

We conclude that the helical field model of \citet{fp00a,fp00c} is
consistent with our observations.  This model predicts the
depolarization along the axis of filaments and the position angle
patterns, as well as explaining the $r^{-2}$ density profile observed
by \citet{joh99}.  Although a quantitative model of OMC-3 as a bent
filament is not yet complete, we are actively pursuing this
possibility as a promising explanation for our data.  There is no
reason to suppose that a filament extending over a parsec should
maintain a single inclination relative to the plane of the sky.
Conversely, the misalignment of vectors toward MMS8 and MMS9 may not
be a misalignment at all if a second filament is juxtaposed on the
\intfil\ at their positions.  The current data set is not extensive
enough to distinguish between these two possible filament-field
geometries.  Polarization mapping to the east, west and south of the
current mapped area would provide more insight into the magnetic field
geometry in this region.  High resolution data are also needed to
further investigate other sources of the depolarization effect along
the axis of the filament.

\acknowledgements

The authors would like to thank J.~Greaves, T.~Jenness, and
G.~Moriarty-Schieven at the JCMT for their assistance with problems
both large and small during and especially after observing.  Thanks to
D.~Johnstone for making the 850 \micron\ scan map of the \intfil\
available to us and to J.~Brown for access to the Faraday rotation
data prior to its publication.  We would also like to thank our
referee for several suggestions which led to improvements in this
manuscript.  The research of BCM and CDW is supported through grants
from the Natural Sciences and Engineering Research Council of Canada.
BCM acknowledges funding from Ontario Graduate Scholarships.  JDF
acknowledges support from a postdoctoral fellowship, jointly funded by
the Natural Sciences and Engineering Council and the Canadian
Institute for Theoretical Astrophysics.

\clearpage

\appendix
\section{Effects of a Polarized Reference (or Sky) Position}
\label{appendixA}

\newcommand{\pobs}{$p_{obs}$}
\newcommand{\paobs}{$\theta_{obs}$}
\newcommand{\Iobs}{$I_{obs}$}
\newcommand{\ps}{$p_s$}
\newcommand{\pas}{$\theta_s$}
\newcommand{\Is}{$I_s$}
\newcommand{\pchop}{$p_r$}
\newcommand{\pachop}{$\theta_r$}
\newcommand{\Ichop}{$I_r$}

The quality of submillimeter polarimetry data can be strongly affected
by chopping on and off the source during observing and sky noise
removal in data reduction.  The former can affect the results if the
difference in intensity of the source minus chop is not actually zero,
or if the chop positions, commonly called the reference positions, are
significantly polarized.  Sky removal requires one to select one or
more SCUBA bolometers which are devoid of significant flux to estimate
sky variations on the timescale of 1 s.  If the flux in these
bolometers is in fact polarized or non-zero, then polarization data
for the whole data set could be affected.

Typically, in reducing polarimetry data, the assumption is made that
fluxes in the chop and sky bolometer positions are negligible and
unpolarized.  These assumptions are most likely valid in the case of
point-like sources (such as protostars) in non-clustered environments
(such as Taurus or Bok globules) where the background, even if
non-zero, is flat enough that chopping reduces the background
effectively to zero.  More typically in star-forming regions, however,
bright cores may be embedded within more extended structures, and even
the largest chop throws of 150-180\arcsec\ (at the JCMT) may be
insufficient to reach ``empty'', unpolarized sky.  If one were only
interested in cores, one could use small chops, removing significant
amounts of extended flux, but even in such cases, the difference
between the on-source and chop positions may not be zero at the edges
of the array given the rapid declines in surface intensities.  For
example, 850 $\mu$m data of the \intfil\ in which OMC-3 is located has
been shown to exhibit a flux profile of $r^{-1}$, which implies a
variation of flux across the filament on the scale of the chop throw
\citep{joh99}.

The effects of a polarized reference, or chop, position on the
observed map are not necessarily intuitive.  In order to illustrate
the behavior of observed polarization percentage and position angle,
we have used a simple model in which the polarization properties of
the source and reference positions are known exactly.  By subtracting
the reference from the source polarization, the behavior of the
observed polarization vectors can be compared to the input source
values.

Linear polarization is defined by the Stokes parameters $Q$ and $U$
(and the total unpolarized intensity $I$),
\begin{equation}
Q = I_p \times \cos (2 \theta) \ \ \ U = I_p \times \sin(2 \theta)
\label{qandu}
\end{equation}

\noindent where $I_p$ is the polarized intensity, the product of the
polarization percentage, $p$, and the total intensity, $I$, and
$\theta$ is the polarization position angle.  Note that under these
definitions, $Q$ and $U$ execute a period in 180$^\circ$ instead of
360$^\circ$.  This reflects the reality that $180^\circ$ offsets are
not detectable in linear polarization.

The percentage polarization is defined by
\begin{equation}
p = \frac{\sqrt{Q^2 + U^2}}{I} \times 100\%
\label{perpoln}
\end{equation}

\noindent while the position angle is given by
\begin{equation}
\theta = \frac{1}{2} \arctan{\left ( \frac{U}{Q} \right )}
\label{posangle}
\end{equation}

\noindent where $U/Q > 0$ for $-180^\circ < 2\theta < -90^\circ$ and $
0^\circ < 2\theta < 90^\circ$.  A ratio of $U/Q <0$ can be found only
where $-90 < 2\theta < 0^\circ$ and $90^\circ < 2\theta < 180^\circ$.
Finally, $U/Q = 0$ only where $2\theta = 0$ and $Q \ne 0$.  Where both
$Q=0$ and $U=0$ observationally, the source must be unpolarized.

If one defines \Is, \ps, and \pas\ values for a source and
\Ichop, \pchop, and \pachop\ for a reference position, then one can
deduce the quantities one would observe: \Iobs, \pobs, and \paobs\ by
simple subtraction of $I$, $Q$, and $U$ values at the two positions.
In this exercise, we do not include an estimate of the rms noise,
which serves primarily to truncate the useful data set observed at low
$I$, $Q$ and $U$ values.  Additionally, although chopping is typically
done to reference positions on either side of the source field, we
will consider only the source data and one reference position.

SCUBA and other bolometric arrays (SHARC, BOLOCAM) sample a wide area
of sky compared to previous single bolometer instruments.  Thus, many
flux levels may be present across their fields of view.  One might
suspect that during SCUBA observations (which allow a maximum chop
throw of 3\arcmin), the reference position could be so close spatially
to the source position that the same polarization properties could be
present at both positions.  Smoothly varying polarization patterns
have been observed in many star-forming regions (e.g.\ see
\citet{dot00}).  Even in such cases, the observed polarization vector
could be adversely affected; the magnitude of the effect depends on
the relative polarized flux (i.e.\ $p \times I$) between the source
and reference fields.

One can consider two straightforward cases which illustrate the
effects of chopping on the observed polarization vectors.  In the
first case, consider a source polarized at $\theta_s =0^\circ$
and a reference polarization of $\theta_r =-90^\circ$.  Using equation
(\ref{qandu}), the values of $Q_s$, $U_s$, $Q_r$, and $U_r$ can be
calculated in terms of the polarized intensities at each position:
$I_{p,s}$ and $I_{p,r}$.  Subtraction of the reference values of $Q$
and $U$ show that the observed quantity, $U_{obs}$, remains zero,
so by equation (\ref{posangle}), the position angle measured will be
the same as that of the source.  However, application of equation
(\ref{perpoln}) yields the result
\begin{displaymath}
p_{obs} = p_s \times \left [ \frac{1 + (I_{p,r}/I_{p,s})}{1 - (I_r/I_s)} \right ]
\end{displaymath}

\noindent where \Iobs\ has been replaced by \Is\ $-$ \Ichop.
Therefore, the true \ps\ is recovered only if both $I_{p,r}/I_{p,s}$
and $I_r/I_s$ approach zero.

In the second case, consider a source polarized at $\theta_s = 0^\circ$
and a reference polarization with position angle $\theta_r = 45^\circ$.
A similar calculation of respective $Q$ and $U$ values, subtraction
and application of equations (\ref{posangle}) and (\ref{perpoln})
reveals that since $U_{obs} \ne 0$, both the position angle and the
percentage polarization will in this case differ from those of the
source.  The observed position angle is a function of $I_{p,r}/I_{p,s}$:
\begin{displaymath}
\theta_{obs} = \frac{1}{2} \arctan{\left ( \frac{-I_{p,r}}{I_{p,s}} \right )}
\end{displaymath}

\noindent while the percentage polarization is again a function of
$I_{p,r}/I_{p,s}$ and $I_r/I_s$:
\begin{displaymath}
p_{obs} = p_s \times \left [\frac{\sqrt{1+(I_{p,r}/I_{p,s})^2}}{1-(I_r/I_s)}\right ].
\end{displaymath}

In order to quantify this effect over many conditions, we subtracted a
reference polarization from a source polarization under several
different cases outlined in Table \ref{tableA1}.  In all cases, a
source polarization of 10\% and a position angle of 0$^\circ$ were
used, with varying ranges of source intensity.\footnote{Calculations
can be done with different \pas\ values, but the results are
completely identical to those presented here.  What is relevant is the
difference between \pas\ and \pachop\, not the absolute value of
either.}  The reference polarization angle is assigned an offset from
this value.  The source total intensity is assigned a range of values
from 1 to 20, and the reference flux is assumed to have a uniform
polarized intensity equal to some fraction that of the source peak.

Figure \ref{polchop_p} plots \pobs\ as a function of \Iobs\ for Cases
A, B and F.  Where the reference polarization is aligned with the
source polarization, the correct \ps\ is measured for all \Iobs $ > 0$
in Cases A and B.  However, in Case F, the \pobs\ is underestimated,
since \pchop\ $>$ \ps.  For large offsets between \pas\ and \pachop,
the polarization percentage could be overestimated, particularly for
low $I_{obs}$ values.  Near the source peak, the polarization observed
converges on the true source polarization value.  The values of
\pobs/\ps\ $> 10$ are, in the case of \ps = 10\%, completely
unphysical and would be disregarded in any data set.  Positive or
negative offsets in position angle between source and reference
produce the same \pobs.  For this reason, only the solutions for
positive offsets have been plotted on Figure \ref{polchop_p}.

Figure \ref{polchop_pa} shows that \pas\ is recovered in all three
cases when the source and reference polarizations are aligned.  When
the source and reference polarizations are not aligned, more
interesting results are obtained.  The largest discrepencies in
\paobs\ occur when the differences between \pas\ and \pachop\
are small.  Figure \ref{polchop_pa} illustrates that, for high values
of \Iobs, the largest errors in \paobs\ from \pas\ are observed for
\pas\ $-$ \pachop $= 45^\circ$, which corresponds to 2(\pas\ $-$
\pachop) of $90^\circ$.  For very low \Iobs, smaller offsets can
produce an even larger error.  As offsets increase toward $90^\circ$,
a 2(\pas\ $-$ \pachop) value of 180$^\circ$ is approached.  Since
linear polarization measurements cannot discriminate between vectors
$180^\circ$ apart, the input \pas\ is recovered.  In Case A, which
most closely parallels the MMS6 field of our data, at least in levels
of intensity between the source and chop position (see $\S$
\ref{obs}), the largest error which can be produced in position angle
is $\pm 10^\circ$, and that is only for the lowest values of \Iobs.
In Case B, which has a source peak flux to reference flux ratio
similar to that of the MMS8-9 region in the OMC-3 data set (based on
the \citet{joh99} scan map data), the most extreme errors in position
angle predicted range from $\pm (20-30)^\circ$ even for very low
intensities, as long as our assumption of similar polarization to the
source holds.  Even though we have noted that the largest errors in
\paobs\ occur for small angles, Figure \ref{polchop_pa} shows that in
Case B, the small angle offsets do not dominate, and the \paobs\ is
within $\pm 20^\circ$ of \pas\ until fluxes are less than 10\% of the
peak \Iobs.  We have truncated our OMC-3 data set such that \Iobs\
$> 0.0006$ (volts), which is approximately 10\% of the faintest peak
in our map (MMS7) with a flux of $0.00554 \pm 0.00002$ (volts).  For
still lower intensities (where \Iobs\ approaches 0), the largest
discrepencies from the input \pas\ are no greater than $\pm 40^\circ$.

In comparison, Case F reveals potential errors of $\pm 90^\circ$ for
low \Iobs $> 0$.  If a true observation were done under these
conditions, polarization vectors would be reliable only to a flux
level about 50\% of the observed peak.  Near the peaks, errors are
only on the order of $< \pm 20^\circ$.  Thus, even in this extreme
case where the \pchop\ $>$ \ps\ and the \Ichop\ is a significant
fraction of \Is, it is unlikely that chopping alone could produce an
alignment of \paobs\ across a SCUBA field of view.  In the idealized
scenario we have discussed, the various fluxes across the SCUBA source
field of view will be affected differently by the polarized flux in
the reference position, creating vectors for which \paobs\ varies
systematically with \Iobs.  OMC-3's southern region contains vectors
which are orthogonal to the filament orientation over the whole SCUBA
field (a range of an order of magnitude in total flux).  The vectors
have a mean of $-70^\circ$ with a range up to $\pm 30^\circ$ from that
value.  There is no systematic variation in position angle with
intensity.  We thus conclude that the simple scenario discussed here
(a constant polarized flux at the reference position) cannot be
responsible for the misalignment between the filament and polarization
vectors.  We cannot rule out a variable polarized flux across the
SCUBA footprint at the reference position, but our examination of the
\citet{joh99} data does not reveal large variations in intensity at
those positions.  If polarization percentage were varying, we have no
means of detecting this variation with our data set.

However, polarization has been detected at 350 \micron\ toward the
northern part of OMC-3 using the Hertz polarimeter at the Caltech
Submillimeter Observatory \citep{dow01}.  These data show alignment
along the filament as observed at 850 \micron\ with SCUBA, although
the 350 \micron\ polarization percentages are $\sim 75$\% those of the
JCMT \citep{hil00}.  Since the CSO data were obtained with a chop
throw of 6\arcmin\ (double that of the JCMT), and the behavior of the
polarization pattern is consistent between the two instruments, this
provides some re-assurance that the SCUBA position angle data have not
been grossly affected by a significant polarized flux at the reference
position.

In practice, efforts should be made to select reference positions
which are devoid of significant flux compared to the flux levels in
the source field of view.  Figures \ref{polchop_p} and
\ref{polchop_pa} show that even if the polarization percentage is
comparable in both fields, the effects on the observable quantities
recovered are minimized greatly if the total flux levels are low at
the reference position.

One of the key ``observable'' relationships in our data set of OMC-3
is the depolarization effect measured along the length of the
filament.  Systematically lower polarization percentages are measured
toward regions of higher intensities.  Figure \ref{logpI_expected}
shows the log \pobs\ versus log ($I_{obs}/I_{obs,peak}$) plots for
offsets of $\pm 90^\circ$ between source and reference polarizations,
which produces the largest error in polarization percentage.  Unlike
the OMC-3 data, for which the log-log plot of Figure \ref{pvsI} looks
reasonably linear (given the noise and scatter), the depolarization
effects of Figure \ref{logpI_expected} clearly are not linear.
However, it is possible to sketch in a maximum and minimum slope.  The
minimum and maximum slopes encompass a range of slopes consistent with
variation of percentage polarization with observed intensity for each
of the six cases.  Table \ref{tableA1} records the slopes; as one
would expect, the depolarization produced is smallest for reference
positions with low values of \Ichop\ and small values of \pchop.

None of the slopes generated for data at high $I_{obs}/I_{obs,peak}$
reflect the slope of $-0.65$ derived for the OMC-3 data set.  The
closest effects are for reference fluxes 25\% of the flux peak, which
we believe does not represent the observed conditions in any part of
OMC-3.  The slopes at high $I_{obs}/I_{obs,peak}$ are the most
likely to be observed in real data, since they are exhibited where
signal-to-noise will be high.  However, at lower
$I_{obs}/I_{obs,peak}$, steeper slopes could be produced, and for
completeness, we have included these as maximum depolarization effects
produced by each case.  For all but the lowest reference flux
considered (2\% of the source peak), a slope of $-0.65$ could be
explained by polarized flux in the reference position.  We note,
however, that these slopes are produced only in regions of low flux,
whereas our data continue to exhibit this slope even at the highest
flux values.  

These calculations reveal that the effective ``depolarization''
created by chopping onto a polarized reference position is dependent
on the intensity and degree of polarization present as well as the
offset between the source and reference position angle.  We conclude
that in the region of MMS8-9, it is possible that the depolarization
observed could result from the scenarios described in Cases B, or
possibly E.  Cases C and F represent an extreme we do not believe
exists in our data set.  However, we note that the northern parts of
OMC-3 have much lower reference fluxes compared to peak source
positions (again, based on the Johnstone \& Bally map).  Near MMS6,
for instance, the scenario is closer to Case A (or maybe Case D), for
which the depolarization effect observed cannot be attributed to
chopping effects.

\section{Sky Subtraction}
\label{appendixB}

Bolometers must be carefully chosen for sky subtraction.  Ideally,
bolometers should be free of emission and, for polarimetry,
unpolarized.  SCUBA has a large field of view ($>2.3$\arcmin) and when
observing point sources, there are typically many ``empty'' bolometers
to chose from for sky subtraction.  Extended sources prove more
difficult.

Figure \ref{appendixB_fig2} compares the data obtained in 6 sets of 3
consecutive observations each toward the MMS8 and MMS9 region.  Taking
3 consecutive scans of the same source allows the resultant maps to be
quickly combined for increased signal-to-noise at the telescope and
minimizes sky rotation between them.  Sets 1 and 2 were obtained on 5
September; Sets 3-5 on 6 September; and Set 6 on 7 September.  Before
sky subtraction is performed, the polarization position angles appear
highly uniform, but are not oriented in the same direction in each
data set (see the left panels of Figure \ref{appendixB_fig2}).  For
example, Set 1 exhibits vectors which align closely to the filament,
while Set 3 appears just the opposite. However, the sky subtracted
versions of each data set reveal very similar polarization patterns,
although these generally appear messier than the unsubtracted data.
Generally, in the sky subtracted maps, the overall orientation of the
vectors is approximately east to west, although there is some high
scatter in the lower signal-to-noise regions.

The uniformity in the non-sky subtracted frames is easily understood
since the dominant factor in the polarization detected is due to sky.
If the opacity of the sky is changing during a single observation
(which it is, necessitating the removal of the sky's effects), then
one can expect a very uniform polarization map.  For instance, if the
sky becomes steadily more opaque during an observation, then maps made
at each successive position of the waveplate will contain fainter
fluxes.  When the waveplate positions are paired up and subtracted to
deduce $Q$ and $U$, then the results must be positive.  If the sky is
dominant, than $Q$ and $U$ will produce approximately the same fluxes,
both positive, which yield a position angle of 45$^\circ$ (east of
north).  The removal of sky effects removes this uniformity and leaves
the more structured polarization of the source itself.  The uniformity
of each subset once the sky has been subtracted is a re-assurance that
the subtraction routine is effectively removing sky variations.

\section{Sub-dividing the OMC-3 data set}
\label{appendixC}

Figure \ref{appendixC_fig} shows two subsets of the OMC-3 data.  The
constraints on the data plotted have been relaxed to reflect the fact
that the noise is larger when only half the exposure time is used.
Thus, instead of plotting values with the uncertainty in polarization
percentage, $dp <1$\%, we have plotted those vectors with
$dp<1.4$\%.  Instead of selecting vectors with $p/dp = \sigma_p >6$, we have
plotted those with $\sigma_p>4.2$.

The data reduction for each subset was performed in the same manner as
for the entire data set as described in $\S$ \ref{obs}.  Despite the
fact that some data are missing from each set due to the removal of
noisy bolometers, the polarization patterns shown in Figure
\ref{appendixC_fig} are very consistent with each other and with that
of Figure \ref{mapnew}.  The alignment between filament and vectors in
the north and misalignment in the south is observed in both maps.
Depolarization toward the filament spine is also observed.  Seventy
percent of the polarization percentages of the 190 vectors in common
between these two sets of data are not significantly different from
one another (i.e.\ $\frac{(p_1 - p_2)}{(dp_1 + dp_2)} < 3$).  The
upper limit on the quantity $p_1 - p_2$ is estimated to be $dp_1 +
dp_2$ because we have reason to suppose these errors will be
correlated.

\section{Polarization Percentages and Position Angles}
\label{appendixD}

Table \ref{allthedata} contains the percentage polarizations and
position angles as plotted in Figure \ref{mapnew}.  The positions are
given as arcsecond offsets from a position near the peak of MMS6, at
J2000 coordinates $\alpha = 05^{\rm h}35^{\rm m}$23\fs5 and $\delta =
-05^\circ 01^\prime$ 32\farcs2 ($\alpha = 05^{\rm h}32^{\rm m}$55\fs6
and $\delta = -05^\circ 03^\prime$ 25\farcs0 in B1950).

\begin{deluxetable}{ccccc}
\tablecolumns{5}
\tablewidth{0pc}
\tablecaption{Observing Parameters for Jiggle Mapping}
\tablehead{
\multicolumn{2}{c}{Pointing Center} & \colhead{Chop Throw} & \colhead{Chop Position Angle} & \colhead{Number of} \\ 
\colhead{R.A. (J2000)} & \colhead{Dec. (J2000)} & \colhead{(\arcsec)} & \colhead{(east of north)} & \colhead{Times Observed} }
\startdata
$05^{\rm h}35^{\rm m}$19\fs3 & $-$05\degr00\arcmin36\farcs9 & 150 & 30\degr\ & 6 \\
$05^{\rm h}35^{\rm m}$18\fs2 & $-$05\degr00\arcmin21\farcs8 & 150 & 30\degr\ & 17 \\
$05^{\rm h}35^{\rm m}$23\fs5 & $-$05\degr01\arcmin32\farcs2 & 150 & 100\degr\ & 6 \\
$05^{\rm h}35^{\rm m}$26\fs5 & $-$05\degr03\arcmin57\farcs4 & 150 & 100\degr\ & 9 \\
$05^{\rm h}35^{\rm m}$27\fs5 & $-$05\degr03\arcmin32\farcs5 & 150 & 90\degr\ & 9 \\
$05^{\rm h}35^{\rm m}$26\fs5 & $-$05\degr05\arcmin31\farcs4 & 150 & 65\degr\ & 8 \\
$05^{\rm h}35^{\rm m}$27\fs5 & $-$05\degr05\arcmin21\farcs5 & 150 & 65\degr\ & 14 \\
\enddata
\label{observing_parameters}
\end{deluxetable}

\begin{deluxetable}{crccc}
\tablecolumns{5} 
\tablewidth{0pc} 
\tablecaption{Systematic Depolarization Created by Chopping onto Polarized Sky} 
\tablehead{
\colhead{} & \colhead{} & \colhead{} & \multicolumn{2}{c}{
Slope of log \pobs\ vs.~log \Iobs\ at \pas\ $-$ \pachop\ = 90\degr} \\
\colhead{Case} & \colhead{\Ichop} & \colhead{\pchop} &
\colhead{minimum (high \Iobs)} & \colhead{maximum (low \Iobs)} }
\startdata
A & 0.4 (2\% source peak) & 10\% & $-0.085$ & $-0.39$ \\
B & 2 (10\% source peak) & 10\% & $-0.28$ & $-0.65$ \\
C & 5 (25\% source peak) & 10\% & $-0.50$ & $-0.81$ \\
D & 0.4 (2\% source peak) & 20\% & $-0.085$ & $-0.49$ \\
E & 2 (10\% source peak) & 20\% & $-0.34$ & $-0.76$ \\
F & 5 (25\% source peak) & 20\% & $-0.56$ & $-0.88$ \\
\enddata
\label{tableA1}
\end{deluxetable}

\begin{deluxetable}{rrrrrrr}
\tablecolumns{7}
\tablewidth{0pc}
\tablecaption{OMC-3 850 \micron\ Polarization Data}
\tablehead{
\colhead{$\Delta$ R.A.} & \colhead{$\Delta$ DEC.} & \colhead{$p$} & \colhead{$dp$} & \colhead{$\sigma_p$} & \colhead{$\theta$} & \colhead{$d\theta$} \\
\colhead{(\arcsec)} & \colhead{(\arcsec)} & \colhead{(\%)} & \colhead{(\%)} & \colhead{} & \colhead{($^\circ$)} & \colhead{($^\circ$)}}
\startdata
  $  48.0$ & $-304.5$ &   8.93 &  0.85 &  10.5 & $  88.3$ &  2.7 \\ 
  $  60.0$ & $-292.5$ &  13.03 &  0.78 &  16.6 & $ -78.5$ &  1.7 \\ 
  $  48.0$ & $-292.5$ &  10.83 &  0.47 &  23.0 & $ -86.5$ &  1.2 \\ 
  $  36.0$ & $-292.5$ &  13.92 &  0.61 &  22.7 & $  89.7$ &  1.3 \\ 
  $  60.0$ & $-280.5$ &  12.44 &  0.62 &  20.2 & $ -75.5$ &  1.4 \\ 
  $  48.0$ & $-280.5$ &   5.01 &  0.33 &  15.3 & $ -89.0$ &  1.9 \\ 
  $  36.0$ & $-280.5$ &  10.02 &  0.39 &  25.6 & $ -87.7$ &  1.1 \\ 
  $  24.0$ & $-280.5$ &  13.28 &  0.60 &  22.1 & $  84.7$ &  1.3 \\ 
  $  12.0$ & $-280.5$ &  12.53 &  0.77 &  16.2 & $ -85.3$ &  1.8 \\ 
  $   0.0$ & $-280.5$ &   8.18 &  0.69 &  11.9 & $ -65.9$ &  2.4 \\ 
  $ -12.0$ & $-280.5$ &   7.92 &  0.92 &   8.6 & $ -70.2$ &  3.3 \\ 
  $ 108.0$ & $-268.5$ &   7.36 &  0.39 &  18.8 & $ -73.6$ &  1.5 \\ 
  $  96.0$ & $-268.5$ &   6.94 &  0.43 &  16.1 & $ -65.5$ &  1.8 \\ 
  $  84.0$ & $-268.5$ &  16.76 &  0.75 &  22.2 & $ -71.3$ &  1.3 \\ 
  $  72.0$ & $-268.5$ &   2.66 &  0.62 &   4.3 & $ -68.4$ &  6.7 \\ 
  $  60.0$ & $-268.5$ &   4.33 &  0.36 &  11.9 & $ -66.9$ &  2.4 \\ 
  $  48.0$ & $-268.5$ &   2.80 &  0.18 &  15.5 & $ -79.3$ &  1.8 \\ 
  $  36.0$ & $-268.5$ &   5.73 &  0.25 &  23.0 & $ -79.2$ &  1.2 \\ 
  $  24.0$ & $-268.5$ &   9.01 &  0.40 &  22.6 & $ -85.2$ &  1.3 \\ 
  $  12.0$ & $-268.5$ &   6.22 &  0.42 &  14.8 & $ -87.3$ &  1.9 \\ 
  $   0.0$ & $-268.5$ &   4.63 &  0.54 &   8.6 & $ -79.7$ &  3.3 \\ 
  $ -12.0$ & $-268.5$ &   9.46 &  0.70 &  13.5 & $ -77.9$ &  2.1 \\ 
  $ 120.0$ & $-256.5$ &   2.61 &  0.37 &   7.0 & $ -43.3$ &  4.1 \\ 
  $ 108.0$ & $-256.5$ &   4.93 &  0.29 &  16.8 & $ -54.2$ &  1.7 \\ 
  $  96.0$ & $-256.5$ &   8.87 &  0.41 &  21.8 & $ -62.7$ &  1.3 \\ 
  $  84.0$ & $-256.5$ &  15.87 &  0.77 &  20.7 & $ -78.1$ &  1.4 \\ 
  $  72.0$ & $-256.5$ &   4.84 &  0.63 &   7.7 & $ -69.2$ &  3.7 \\ 
  $  60.0$ & $-256.5$ &   4.49 &  0.25 &  18.3 & $ -71.4$ &  1.6 \\ 
  $  48.0$ & $-256.5$ &   2.60 &  0.15 &  17.5 & $ -68.7$ &  1.6 \\ 
  $  36.0$ & $-256.5$ &   2.09 &  0.19 &  11.1 & $ -71.8$ &  2.6 \\ 
  $  24.0$ & $-256.5$ &   2.00 &  0.34 &   5.8 & $  85.6$ &  4.9 \\ 
  $  12.0$ & $-256.5$ &   8.01 &  0.36 &  22.3 & $ -79.8$ &  1.3 \\ 
  $   0.0$ & $-256.5$ &   7.70 &  0.43 &  17.9 & $ -77.2$ &  1.6 \\ 
  $ -12.0$ & $-256.5$ &   3.80 &  0.73 &   5.2 & $ -67.6$ &  5.5 \\ 
  $ 132.0$ & $-244.5$ &   6.15 &  0.44 &  13.9 & $ -74.3$ &  2.1 \\ 
  $ 120.0$ & $-244.5$ &  10.43 &  0.35 &  29.4 & $ -72.6$ &  1.0 \\ 
  $ 108.0$ & $-244.5$ &   8.13 &  0.43 &  18.9 & $ -73.9$ &  1.5 \\ 
  $  96.0$ & $-244.5$ &  14.15 &  0.64 &  22.1 & $ -81.4$ &  1.3 \\ 
  $  84.0$ & $-244.5$ &  15.15 &  0.76 &  19.8 & $ -85.5$ &  1.4 \\ 
  $  72.0$ & $-244.5$ &  12.76 &  0.47 &  27.4 & $ -77.0$ &  1.0 \\ 
  $  60.0$ & $-244.5$ &   3.29 &  0.20 &  16.8 & $ -63.2$ &  1.7 \\ 
  $  48.0$ & $-244.5$ &   1.59 &  0.11 &  14.3 & $ -60.3$ &  2.0 \\ 
  $  36.0$ & $-244.5$ &   2.18 &  0.15 &  14.7 & $ -59.6$ &  2.0 \\ 
  $  24.0$ & $-244.5$ &   2.93 &  0.29 &  10.2 & $ -70.9$ &  2.8 \\ 
  $  12.0$ & $-244.5$ &   1.41 &  0.30 &   4.6 & $ -62.5$ &  6.2 \\ 
  $   0.0$ & $-244.5$ &  10.02 &  0.42 &  24.0 & $ -80.8$ &  1.2 \\ 
  $ -12.0$ & $-244.5$ &  11.51 &  0.79 &  14.6 & $ -87.1$ &  2.0 \\ 
  $ 132.0$ & $-232.5$ &   7.02 &  0.43 &  16.5 & $ -51.2$ &  1.7 \\ 
  $ 120.0$ & $-232.5$ &   8.95 &  0.43 &  20.8 & $ -66.3$ &  1.4 \\ 
  $ 108.0$ & $-232.5$ &   6.29 &  0.57 &  11.1 & $ -64.4$ &  2.6 \\ 
  $  96.0$ & $-232.5$ &  13.09 &  0.67 &  19.5 & $ -69.2$ &  1.5 \\ 
  $  84.0$ & $-232.5$ &  10.59 &  0.57 &  18.6 & $ -63.6$ &  1.5 \\ 
  $  72.0$ & $-232.5$ &   6.77 &  0.35 &  19.1 & $ -72.7$ &  1.5 \\ 
  $  60.0$ & $-232.5$ &   3.71 &  0.18 &  20.6 & $ -60.1$ &  1.4 \\ 
  $  48.0$ & $-232.5$ &   2.58 &  0.10 &  26.3 & $ -64.3$ &  1.1 \\ 
  $  36.0$ & $-232.5$ &   3.02 &  0.13 &  22.4 & $ -68.9$ &  1.3 \\ 
  $  24.0$ & $-232.5$ &   4.41 &  0.24 &  18.1 & $ -68.4$ &  1.6 \\ 
  $  12.0$ & $-232.5$ &   5.78 &  0.29 &  20.1 & $ -60.1$ &  1.4 \\ 
  $   0.0$ & $-232.5$ &   4.01 &  0.42 &   9.7 & $ -88.2$ &  3.0 \\ 
  $ -12.0$ & $-232.5$ &   6.22 &  0.61 &  10.1 & $ -48.0$ &  2.8 \\ 
  $ -24.0$ & $-232.5$ &  10.85 &  0.83 &  13.0 & $ -51.5$ &  2.2 \\ 
  $ 132.0$ & $-220.5$ &   6.35 &  0.77 &   8.2 & $ -48.8$ &  3.5 \\ 
  $ 120.0$ & $-220.5$ &   6.95 &  0.80 &   8.7 & $ -81.0$ &  3.3 \\ 
  $ 108.0$ & $-220.5$ &   4.97 &  0.71 &   7.0 & $ -53.9$ &  4.1 \\ 
  $  96.0$ & $-220.5$ &   9.44 &  0.72 &  13.0 & $ -59.0$ &  2.2 \\ 
  $  84.0$ & $-220.5$ &   4.64 &  0.46 &  10.0 & $ -74.2$ &  2.9 \\ 
  $  72.0$ & $-220.5$ &   4.90 &  0.30 &  16.4 & $ -63.6$ &  1.7 \\ 
  $  60.0$ & $-220.5$ &   4.47 &  0.17 &  26.4 & $ -64.1$ &  1.1 \\ 
  $  48.0$ & $-220.5$ &   1.50 &  0.09 &  16.2 & $ -61.4$ &  1.8 \\ 
  $  36.0$ & $-220.5$ &   2.82 &  0.15 &  18.7 & $ -64.6$ &  1.5 \\ 
  $  24.0$ & $-220.5$ &   4.44 &  0.39 &  11.4 & $ -53.1$ &  2.5 \\ 
  $  12.0$ & $-220.5$ &   5.70 &  0.55 &  10.4 & $ -68.4$ &  2.8 \\ 
  $   0.0$ & $-220.5$ &   4.87 &  0.40 &  12.1 & $ -76.0$ &  2.4 \\ 
  $ -12.0$ & $-220.5$ &   7.13 &  0.48 &  14.9 & $ -72.8$ &  1.9 \\ 
  $ -24.0$ & $-220.5$ &   8.37 &  0.76 &  10.9 & $ -59.8$ &  2.6 \\ 
  $  96.0$ & $-208.5$ &   3.18 &  0.90 &   3.5 & $ -55.8$ &  8.1 \\ 
  $  84.0$ & $-208.5$ &   6.67 &  0.46 &  14.6 & $ -62.7$ &  2.0 \\ 
  $  72.0$ & $-208.5$ &   4.09 &  0.25 &  16.1 & $ -75.2$ &  1.8 \\ 
  $  60.0$ & $-208.5$ &   2.61 &  0.15 &  17.3 & $ -63.5$ &  1.7 \\ 
  $  48.0$ & $-208.5$ &   2.42 &  0.12 &  20.3 & $ -64.2$ &  1.4 \\ 
  $  36.0$ & $-208.5$ &   4.46 &  0.22 &  20.3 & $ -68.5$ &  1.4 \\ 
  $  24.0$ & $-208.5$ &  10.29 &  0.57 &  18.2 & $ -66.8$ &  1.6 \\ 
  $  12.0$ & $-208.5$ &   5.06 &  0.82 &   6.2 & $ -73.2$ &  4.6 \\ 
  $   0.0$ & $-208.5$ &   8.26 &  0.44 &  18.6 & $ -61.5$ &  1.5 \\ 
  $ -12.0$ & $-208.5$ &  10.23 &  0.55 &  18.6 & $ -61.2$ &  1.5 \\ 
  $  96.0$ & $-196.5$ &   8.72 &  0.79 &  11.1 & $ -52.5$ &  2.6 \\ 
  $  84.0$ & $-196.5$ &   4.97 &  0.62 &   8.0 & $ -72.9$ &  3.6 \\ 
  $  72.0$ & $-196.5$ &   3.25 &  0.33 &   9.7 & $ -72.5$ &  2.9 \\ 
  $  60.0$ & $-196.5$ &   2.53 &  0.22 &  11.5 & $ -57.6$ &  2.5 \\ 
  $  48.0$ & $-196.5$ &   2.88 &  0.23 &  12.4 & $ -59.8$ &  2.3 \\ 
  $  36.0$ & $-196.5$ &   5.01 &  0.46 &  10.9 & $ -81.3$ &  2.6 \\ 
  $   0.0$ & $-196.5$ &  16.91 &  0.68 &  24.8 & $ -70.5$ &  1.2 \\ 
  $ -12.0$ & $-196.5$ &  15.73 &  0.73 &  21.6 & $ -57.5$ &  1.3 \\ 
  $  72.0$ & $-184.5$ &   8.80 &  0.92 &   9.6 & $ -79.4$ &  3.0 \\ 
  $  60.0$ & $-184.5$ &   3.46 &  0.43 &   8.0 & $ -71.8$ &  3.6 \\ 
  $  48.0$ & $-184.5$ &   6.78 &  0.47 &  14.4 & $ -66.6$ &  2.0 \\ 
  $  36.0$ & $-184.5$ &  14.11 &  0.92 &  15.3 & $ -79.6$ &  1.9 \\ 
  $  60.0$ & $-172.5$ &   8.09 &  0.55 &  14.8 & $ -71.8$ &  1.9 \\ 
  $  48.0$ & $-172.5$ &   4.89 &  0.43 &  11.4 & $ -50.7$ &  2.5 \\ 
  $  60.0$ & $-160.5$ &   8.68 &  0.47 &  18.4 & $ -56.6$ &  1.6 \\ 
  $  48.0$ & $-160.5$ &   5.51 &  0.29 &  19.3 & $ -60.4$ &  1.5 \\ 
  $  36.0$ & $-160.5$ &   4.77 &  0.41 &  11.7 & $ -53.5$ &  2.4 \\ 
  $ 108.0$ & $-148.5$ &   5.74 &  0.92 &   6.2 & $  22.3$ &  4.6 \\ 
  $  60.0$ & $-148.5$ &   3.21 &  0.36 &   9.0 & $ -16.2$ &  3.2 \\ 
  $  48.0$ & $-148.5$ &   2.03 &  0.17 &  11.8 & $ -21.4$ &  2.4 \\ 
  $  36.0$ & $-148.5$ &   2.94 &  0.25 &  11.6 & $ -38.5$ &  2.5 \\ 
  $  24.0$ & $-148.5$ &   2.32 &  0.76 &   3.1 & $ -41.5$ &  9.4 \\ 
  $ 108.0$ & $-136.5$ &   2.33 &  0.78 &   3.0 & $ -55.8$ &  9.6 \\ 
  $  96.0$ & $-136.5$ &   5.13 &  0.69 &   7.5 & $ -37.7$ &  3.8 \\ 
  $  84.0$ & $-136.5$ &   5.02 &  0.65 &   7.8 & $ -44.2$ &  3.7 \\ 
  $  72.0$ & $-136.5$ &   2.24 &  0.40 &   5.5 & $ -25.6$ &  5.2 \\ 
  $  24.0$ & $-136.5$ &   3.18 &  0.46 &   7.0 & $ -37.7$ &  4.1 \\ 
  $  96.0$ & $-124.5$ &   4.55 &  0.87 &   5.2 & $  -5.7$ &  5.5 \\ 
  $  84.0$ & $-124.5$ &   1.88 &  0.53 &   3.5 & $ -25.3$ &  8.1 \\ 
  $  72.0$ & $-124.5$ &   2.16 &  0.27 &   8.0 & $   0.2$ &  3.6 \\ 
  $  60.0$ & $-124.5$ &   2.85 &  0.17 &  16.5 & $  -6.4$ &  1.7 \\ 
  $  48.0$ & $-124.5$ &   2.01 &  0.14 &  14.0 & $ -14.8$ &  2.0 \\ 
  $  36.0$ & $-124.5$ &   1.69 &  0.19 &   8.8 & $ -17.9$ &  3.3 \\ 
  $  12.0$ & $-124.5$ &   7.35 &  0.56 &  13.2 & $  -9.2$ &  2.2 \\ 
  $  96.0$ & $-112.5$ &   4.39 &  0.92 &   4.8 & $ -49.0$ &  6.0 \\ 
  $  84.0$ & $-112.5$ &   4.15 &  0.58 &   7.2 & $ -25.6$ &  4.0 \\ 
  $  72.0$ & $-112.5$ &   2.75 &  0.31 &   8.9 & $  -6.3$ &  3.2 \\ 
  $  60.0$ & $-112.5$ &   1.81 &  0.24 &   7.5 & $  -2.8$ &  3.8 \\ 
  $  48.0$ & $-112.5$ &   1.77 &  0.20 &   9.0 & $ -17.4$ &  3.2 \\ 
  $  36.0$ & $-112.5$ &   1.77 &  0.25 &   7.1 & $ -19.9$ &  4.1 \\ 
  $  24.0$ & $-112.5$ &   1.49 &  0.50 &   3.0 & $ -76.6$ &  9.6 \\ 
  $  12.0$ & $-112.5$ &   2.53 &  0.49 &   5.1 & $ -35.8$ &  5.6 \\ 
  $   0.0$ & $-112.5$ &   4.34 &  0.74 &   5.9 & $  20.1$ &  4.9 \\ 
  $  96.0$ & $-100.5$ &   3.31 &  0.85 &   3.9 & $  50.8$ &  7.4 \\ 
  $  84.0$ & $-100.5$ &   4.19 &  0.70 &   6.0 & $  16.8$ &  4.8 \\ 
  $  72.0$ & $-100.5$ &   8.22 &  0.63 &  13.1 & $  12.9$ &  2.2 \\ 
  $  48.0$ & $-100.5$ &   3.23 &  0.30 &  10.8 & $ -34.7$ &  2.7 \\ 
  $  36.0$ & $-100.5$ &   1.95 &  0.32 &   6.1 & $ -26.8$ &  4.7 \\ 
  $  24.0$ & $-100.5$ &   3.65 &  0.79 &   4.6 & $  58.0$ &  6.2 \\ 
  $  12.0$ & $-100.5$ &  11.23 &  0.95 &  11.9 & $ -71.7$ &  2.4 \\ 
  $  48.0$ & $ -88.5$ &   2.14 &  0.49 &   4.3 & $ -15.7$ &  6.6 \\ 
  $  36.0$ & $ -88.5$ &   1.98 &  0.32 &   6.2 & $ -45.7$ &  4.6 \\ 
  $  24.0$ & $ -88.5$ &   3.78 &  0.45 &   8.4 & $ -16.5$ &  3.4 \\ 
  $  12.0$ & $ -88.5$ &   3.22 &  0.75 &   4.3 & $ -53.0$ &  6.6 \\ 
  $  48.0$ & $ -76.5$ &   7.98 &  0.64 &  12.4 & $ -31.6$ &  2.3 \\ 
  $  36.0$ & $ -76.5$ &   4.00 &  0.34 &  11.9 & $ -35.2$ &  2.4 \\ 
  $  12.0$ & $ -76.5$ &   5.99 &  0.76 &   7.9 & $ -13.5$ &  3.6 \\ 
  $  48.0$ & $ -64.5$ &   7.59 &  0.86 &   8.9 & $  -2.7$ &  3.2 \\ 
  $  36.0$ & $ -64.5$ &   2.00 &  0.33 &   6.1 & $  23.3$ &  4.7 \\ 
  $  24.0$ & $ -64.5$ &   2.35 &  0.33 &   7.1 & $  30.1$ &  4.0 \\ 
  $  12.0$ & $ -64.5$ &   4.88 &  0.65 &   7.5 & $  -0.5$ &  3.8 \\ 
  $  12.0$ & $ -52.5$ &   5.53 &  0.52 &  10.6 & $   4.2$ &  2.7 \\ 
  $  36.0$ & $ -40.5$ &   7.18 &  0.69 &  10.4 & $   0.5$ &  2.8 \\ 
  $  24.0$ & $ -40.5$ &   2.86 &  0.48 &   6.0 & $  39.3$ &  4.8 \\ 
  $  36.0$ & $ -28.5$ &   5.59 &  0.81 &   6.9 & $   3.1$ &  4.1 \\ 
  $  24.0$ & $ -28.5$ &   2.23 &  0.52 &   4.3 & $ -40.8$ &  6.7 \\ 
  $  12.0$ & $ -28.5$ &   1.00 &  0.32 &   3.1 & $ -32.1$ &  9.2 \\ 
  $   0.0$ & $ -28.5$ &   3.46 &  0.37 &   9.3 & $ -41.3$ &  3.1 \\ 
  $ -12.0$ & $ -28.5$ &   9.02 &  0.73 &  12.3 & $ -44.9$ &  2.3 \\ 
  $ -48.0$ & $ -28.5$ &   8.33 &  0.62 &  13.4 & $ -37.0$ &  2.1 \\ 
  $ -60.0$ & $ -28.5$ &   8.43 &  0.68 &  12.3 & $ -20.9$ &  2.3 \\ 
  $  36.0$ & $ -16.5$ &   6.90 &  0.66 &  10.5 & $  24.7$ &  2.7 \\ 
  $  24.0$ & $ -16.5$ &   2.43 &  0.50 &   4.8 & $   9.2$ &  5.9 \\ 
  $   0.0$ & $ -16.5$ &   1.71 &  0.23 &   7.5 & $ -26.5$ &  3.8 \\ 
  $ -12.0$ & $ -16.5$ &   4.08 &  0.37 &  10.9 & $   3.5$ &  2.6 \\ 
  $ -24.0$ & $ -16.5$ &   4.40 &  0.88 &   5.0 & $  -0.4$ &  5.7 \\ 
  $ -36.0$ & $ -16.5$ &   3.93 &  0.71 &   5.5 & $ -41.2$ &  5.2 \\ 
  $ -48.0$ & $ -16.5$ &   1.48 &  0.36 &   4.2 & $ -33.0$ &  6.9 \\ 
  $ -72.0$ & $ -16.5$ &   5.11 &  0.48 &  10.7 & $  14.1$ &  2.7 \\ 
  $ -84.0$ & $ -16.5$ &   8.94 &  0.81 &  11.0 & $  39.5$ &  2.6 \\ 
  $  12.0$ & $  -4.5$ &   1.72 &  0.27 &   6.3 & $ -33.8$ &  4.5 \\ 
  $   0.0$ & $  -4.5$ &   1.81 &  0.16 &  11.4 & $ -41.8$ &  2.5 \\ 
  $ -12.0$ & $  -4.5$ &   1.57 &  0.28 &   5.5 & $ -10.1$ &  5.2 \\ 
  $ -24.0$ & $  -4.5$ &   4.00 &  0.32 &  12.5 & $ -19.9$ &  2.3 \\ 
  $ -36.0$ & $  -4.5$ &   2.77 &  0.33 &   8.5 & $ -26.7$ &  3.4 \\ 
  $ -48.0$ & $  -4.5$ &   1.23 &  0.25 &   5.0 & $ -19.0$ &  5.7 \\ 
  $ -60.0$ & $  -4.5$ &   1.42 &  0.32 &   4.4 & $  -5.3$ &  6.5 \\ 
  $ -72.0$ & $  -4.5$ &   3.13 &  0.28 &  11.4 & $  14.4$ &  2.5 \\ 
  $ -84.0$ & $  -4.5$ &   2.23 &  0.39 &   5.8 & $ -17.2$ &  5.0 \\ 
  $ -96.0$ & $  -4.5$ &  11.17 &  0.77 &  14.5 & $  13.8$ &  2.0 \\ 
  $  36.0$ & $   7.5$ &   5.73 &  0.98 &   5.9 & $  33.0$ &  4.9 \\ 
  $  24.0$ & $   7.5$ &   2.46 &  0.62 &   4.0 & $  41.0$ &  7.2 \\ 
  $  12.0$ & $   7.5$ &   2.67 &  0.32 &   8.3 & $ -24.1$ &  3.5 \\ 
  $   0.0$ & $   7.5$ &   1.11 &  0.12 &   9.0 & $ -35.6$ &  3.2 \\ 
  $ -12.0$ & $   7.5$ &   2.75 &  0.12 &  23.1 & $ -31.3$ &  1.2 \\ 
  $ -24.0$ & $   7.5$ &   3.75 &  0.19 &  19.7 & $ -28.5$ &  1.5 \\ 
  $ -36.0$ & $   7.5$ &   3.12 &  0.22 &  13.9 & $ -30.9$ &  2.1 \\ 
  $ -48.0$ & $   7.5$ &   2.59 &  0.23 &  11.3 & $ -45.8$ &  2.5 \\ 
  $ -60.0$ & $   7.5$ &   3.41 &  0.29 &  11.7 & $  -3.7$ &  2.4 \\ 
  $ -72.0$ & $   7.5$ &   2.28 &  0.28 &   8.1 & $  16.8$ &  3.5 \\ 
  $ -84.0$ & $   7.5$ &   3.50 &  0.38 &   9.3 & $ -21.0$ &  3.1 \\ 
  $ -96.0$ & $   7.5$ &   3.96 &  0.53 &   7.5 & $  -1.7$ &  3.8 \\ 
  $  12.0$ & $  19.5$ &   5.13 &  0.39 &  13.3 & $   0.8$ &  2.2 \\ 
  $   0.0$ & $  19.5$ &   4.05 &  0.17 &  23.5 & $ -29.3$ &  1.2 \\ 
  $ -12.0$ & $  19.5$ &   3.63 &  0.10 &  37.1 & $ -32.7$ &  0.8 \\ 
  $ -24.0$ & $  19.5$ &   4.14 &  0.11 &  39.1 & $ -42.1$ &  0.7 \\ 
  $ -36.0$ & $  19.5$ &   3.72 &  0.17 &  21.6 & $ -45.1$ &  1.3 \\ 
  $ -48.0$ & $  19.5$ &   3.40 &  0.18 &  18.9 & $ -60.3$ &  1.5 \\ 
  $ -60.0$ & $  19.5$ &   1.45 &  0.21 &   6.8 & $ -64.8$ &  4.2 \\ 
  $ -72.0$ & $  19.5$ &   2.34 &  0.25 &   9.3 & $ -17.7$ &  3.1 \\ 
  $ -84.0$ & $  19.5$ &   3.68 &  0.38 &   9.6 & $  -0.3$ &  3.0 \\ 
  $ -96.0$ & $  19.5$ &   3.17 &  0.43 &   7.3 & $ -23.5$ &  3.9 \\ 
  $-108.0$ & $  19.5$ &   6.48 &  0.63 &  10.3 & $ -14.7$ &  2.8 \\ 
  $  12.0$ & $  31.5$ &   6.73 &  0.85 &   8.0 & $   4.5$ &  3.6 \\ 
  $   0.0$ & $  31.5$ &   7.10 &  0.30 &  23.3 & $ -18.0$ &  1.2 \\ 
  $ -12.0$ & $  31.5$ &   5.05 &  0.13 &  39.8 & $ -31.0$ &  0.7 \\ 
  $ -24.0$ & $  31.5$ &   2.72 &  0.10 &  27.0 & $ -40.7$ &  1.1 \\ 
  $ -36.0$ & $  31.5$ &   2.01 &  0.08 &  23.7 & $ -50.6$ &  1.2 \\ 
  $ -48.0$ & $  31.5$ &   1.58 &  0.09 &  18.1 & $ -51.7$ &  1.6 \\ 
  $ -72.0$ & $  31.5$ &   1.83 &  0.17 &  10.8 & $ -24.8$ &  2.6 \\ 
  $ -84.0$ & $  31.5$ &   2.39 &  0.25 &   9.5 & $ -25.3$ &  3.0 \\ 
  $ -96.0$ & $  31.5$ &   5.13 &  0.35 &  14.8 & $ -28.6$ &  1.9 \\ 
  $-108.0$ & $  31.5$ &   7.79 &  0.52 &  14.9 & $ -18.6$ &  1.9 \\ 
  $-120.0$ & $  31.5$ &   8.56 &  0.72 &  11.9 & $ -12.9$ &  2.4 \\ 
  $   0.0$ & $  43.5$ &  12.83 &  0.57 &  22.6 & $  -8.7$ &  1.3 \\ 
  $ -12.0$ & $  43.5$ &   5.16 &  0.23 &  22.4 & $ -29.7$ &  1.3 \\ 
  $ -24.0$ & $  43.5$ &   3.39 &  0.15 &  22.1 & $ -46.3$ &  1.3 \\ 
  $ -36.0$ & $  43.5$ &   2.04 &  0.09 &  23.3 & $ -51.8$ &  1.2 \\ 
  $ -48.0$ & $  43.5$ &   1.82 &  0.07 &  27.5 & $ -54.1$ &  1.0 \\ 
  $ -60.0$ & $  43.5$ &   1.88 &  0.07 &  28.4 & $ -45.8$ &  1.0 \\ 
  $ -72.0$ & $  43.5$ &   2.16 &  0.09 &  24.0 & $ -33.0$ &  1.2 \\ 
  $ -84.0$ & $  43.5$ &   3.05 &  0.13 &  23.0 & $ -34.4$ &  1.2 \\ 
  $ -96.0$ & $  43.5$ &   2.05 &  0.24 &   8.5 & $ -32.6$ &  3.4 \\ 
  $-108.0$ & $  43.5$ &   4.89 &  0.38 &  12.8 & $ -85.7$ &  2.2 \\ 
  $-120.0$ & $  43.5$ &   6.92 &  0.47 &  14.9 & $ -24.1$ &  1.9 \\ 
  $-132.0$ & $  43.5$ &   2.88 &  0.66 &   4.4 & $ -43.2$ &  6.5 \\ 
  $ -12.0$ & $  55.5$ &   5.51 &  0.43 &  12.9 & $ -21.2$ &  2.2 \\ 
  $ -24.0$ & $  55.5$ &   2.54 &  0.21 &  12.0 & $ -54.6$ &  2.4 \\ 
  $ -36.0$ & $  55.5$ &   2.45 &  0.13 &  18.5 & $ -44.0$ &  1.5 \\ 
  $ -48.0$ & $  55.5$ &   2.12 &  0.09 &  23.5 & $ -48.4$ &  1.2 \\ 
  $ -60.0$ & $  55.5$ &   2.15 &  0.07 &  29.0 & $ -46.5$ &  1.0 \\ 
  $ -72.0$ & $  55.5$ &   2.32 &  0.07 &  33.7 & $ -43.9$ &  0.8 \\ 
  $ -84.0$ & $  55.5$ &   3.06 &  0.10 &  31.2 & $ -38.8$ &  0.9 \\ 
  $ -96.0$ & $  55.5$ &   2.12 &  0.15 &  14.5 & $ -25.0$ &  2.0 \\ 
  $-108.0$ & $  55.5$ &   2.66 &  0.30 &   8.8 & $ -27.9$ &  3.3 \\ 
  $-120.0$ & $  55.5$ &   3.34 &  0.34 &   9.7 & $ -38.8$ &  2.9 \\ 
  $-132.0$ & $  55.5$ &   2.75 &  0.52 &   5.3 & $ -17.8$ &  5.4 \\ 
  $ -24.0$ & $  67.5$ &   2.75 &  0.38 &   7.2 & $ -52.8$ &  4.0 \\ 
  $ -36.0$ & $  67.5$ &   2.52 &  0.23 &  11.1 & $ -57.2$ &  2.6 \\ 
  $ -48.0$ & $  67.5$ &   2.75 &  0.14 &  19.2 & $ -44.6$ &  1.5 \\ 
  $ -60.0$ & $  67.5$ &   2.85 &  0.10 &  28.3 & $ -51.0$ &  1.0 \\ 
  $ -72.0$ & $  67.5$ &   2.60 &  0.06 &  40.9 & $ -46.3$ &  0.7 \\ 
  $ -84.0$ & $  67.5$ &   2.32 &  0.08 &  29.2 & $ -34.5$ &  1.0 \\ 
  $ -96.0$ & $  67.5$ &   2.72 &  0.13 &  20.5 & $ -45.7$ &  1.4 \\ 
  $-108.0$ & $  67.5$ &   2.36 &  0.24 &   9.9 & $ -18.2$ &  2.9 \\ 
  $-120.0$ & $  67.5$ &   2.45 &  0.37 &   6.7 & $ -89.1$ &  4.3 \\ 
  $-132.0$ & $  67.5$ &   3.01 &  0.44 &   6.9 & $   4.1$ &  4.2 \\ 
  $-144.0$ & $  67.5$ &   4.33 &  0.56 &   7.8 & $ -15.2$ &  3.7 \\ 
  $ -36.0$ & $  79.5$ &   3.78 &  0.30 &  12.6 & $ -43.4$ &  2.3 \\ 
  $ -48.0$ & $  79.5$ &   3.23 &  0.20 &  16.3 & $ -49.9$ &  1.8 \\ 
  $ -60.0$ & $  79.5$ &   3.70 &  0.15 &  24.5 & $ -54.7$ &  1.2 \\ 
  $ -72.0$ & $  79.5$ &   3.54 &  0.10 &  35.2 & $ -52.4$ &  0.8 \\ 
  $ -84.0$ & $  79.5$ &   2.31 &  0.11 &  21.8 & $ -55.3$ &  1.3 \\ 
  $ -96.0$ & $  79.5$ &   2.50 &  0.14 &  17.8 & $ -50.9$ &  1.6 \\ 
  $-108.0$ & $  79.5$ &   1.48 &  0.22 &   6.6 & $ -27.8$ &  4.3 \\ 
  $-120.0$ & $  79.5$ &   3.18 &  0.33 &   9.6 & $  13.9$ &  3.0 \\ 
  $-132.0$ & $  79.5$ &   3.52 &  0.36 &   9.7 & $ -19.5$ &  2.9 \\ 
  $-144.0$ & $  79.5$ &   4.01 &  0.59 &   6.8 & $ -17.6$ &  4.2 \\ 
  $ -36.0$ & $  91.5$ &   7.85 &  0.55 &  14.3 & $ -33.9$ &  2.0 \\ 
  $ -48.0$ & $  91.5$ &   5.37 &  0.35 &  15.5 & $ -53.0$ &  1.8 \\ 
  $ -60.0$ & $  91.5$ &   5.59 &  0.21 &  26.4 & $ -52.3$ &  1.1 \\ 
  $ -72.0$ & $  91.5$ &   5.22 &  0.20 &  26.6 & $ -51.5$ &  1.1 \\ 
  $ -84.0$ & $  91.5$ &   5.14 &  0.17 &  30.8 & $ -47.1$ &  0.9 \\ 
  $ -96.0$ & $  91.5$ &   4.52 &  0.23 &  19.8 & $ -49.6$ &  1.4 \\ 
  $-108.0$ & $  91.5$ &   1.87 &  0.25 &   7.3 & $ -38.3$ &  3.9 \\ 
  $-120.0$ & $  91.5$ &   2.93 &  0.25 &  11.9 & $ -23.9$ &  2.4 \\ 
  $-132.0$ & $  91.5$ &   3.30 &  0.26 &  12.6 & $  -6.4$ &  2.3 \\ 
  $-144.0$ & $  91.5$ &   7.69 &  0.60 &  12.9 & $ -10.0$ &  2.2 \\ 
  $ -48.0$ & $ 103.5$ &  12.31 &  0.98 &  12.5 & $ -68.5$ &  2.3 \\ 
  $ -60.0$ & $ 103.5$ &   6.95 &  0.45 &  15.4 & $ -46.7$ &  1.9 \\ 
  $ -72.0$ & $ 103.5$ &   6.64 &  0.29 &  22.8 & $ -45.1$ &  1.3 \\ 
  $ -84.0$ & $ 103.5$ &   5.15 &  0.27 &  19.1 & $ -62.3$ &  1.5 \\ 
  $ -96.0$ & $ 103.5$ &   3.50 &  0.38 &   9.3 & $ -36.7$ &  3.1 \\ 
  $-108.0$ & $ 103.5$ &   2.50 &  0.33 &   7.6 & $ -18.5$ &  3.8 \\ 
  $-120.0$ & $ 103.5$ &   1.97 &  0.22 &   9.1 & $ -38.6$ &  3.2 \\ 
  $-132.0$ & $ 103.5$ &   3.55 &  0.21 &  16.8 & $ -24.7$ &  1.7 \\ 
  $-144.0$ & $ 103.5$ &   5.90 &  0.59 &  10.0 & $ -16.9$ &  2.9 \\ 
  $ -60.0$ & $ 115.5$ &   4.20 &  0.77 &   5.5 & $ -42.0$ &  5.3 \\ 
  $ -72.0$ & $ 115.5$ &   6.35 &  0.44 &  14.5 & $ -48.1$ &  2.0 \\ 
  $ -84.0$ & $ 115.5$ &   6.91 &  0.39 &  17.9 & $ -52.8$ &  1.6 \\ 
  $ -96.0$ & $ 115.5$ &   4.45 &  0.55 &   8.0 & $ -38.5$ &  3.6 \\ 
  $-108.0$ & $ 115.5$ &   2.98 &  0.47 &   6.4 & $ -50.6$ &  4.5 \\ 
  $-120.0$ & $ 115.5$ &   3.90 &  0.22 &  17.8 & $ -27.3$ &  1.6 \\ 
  $-132.0$ & $ 115.5$ &   4.37 &  0.28 &  15.4 & $  -9.0$ &  1.9 \\ 
  $ -72.0$ & $ 127.5$ &   6.26 &  0.72 &   8.6 & $ -45.4$ &  3.3 \\ 
  $ -84.0$ & $ 127.5$ &   8.35 &  0.55 &  15.3 & $ -43.4$ &  1.9 \\ 
  $ -96.0$ & $ 127.5$ &   4.39 &  0.53 &   8.3 & $ -10.6$ &  3.5 \\ 
  $-108.0$ & $ 127.5$ &   1.85 &  0.50 &   3.7 & $ -19.9$ &  7.8 \\ 
  $-120.0$ & $ 127.5$ &   5.48 &  0.42 &  13.0 & $ -46.6$ &  2.2 \\ 
  $-132.0$ & $ 127.5$ &   5.77 &  0.80 &   7.2 & $ -61.8$ &  4.0 \\ 
  $ -96.0$ & $ 139.5$ &   3.82 &  0.89 &   4.3 & $   3.6$ &  6.7 \\ 
\enddata
\label{allthedata}
\end{deluxetable}

\figcaption[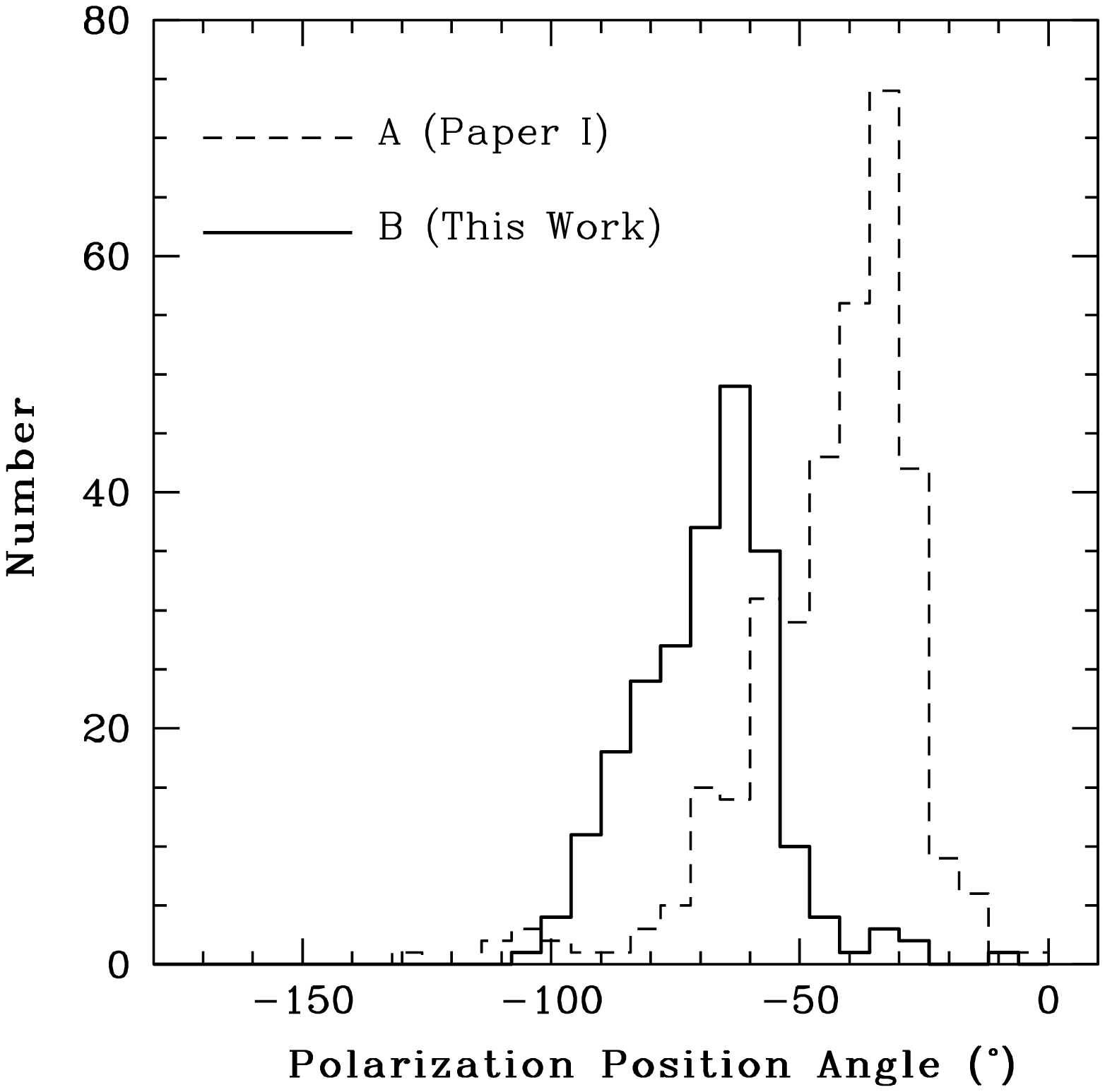]{Comparison of position angle distributions in the
MMS8-9 region.  We show the distribution of position angles for the
southern part of the OMC-3 map as presented in Paper I (distribution
A) and this work (distribution B).  Distribution A can be fit by two
Gaussians of means $-33^\circ$ and $-47^\circ$ (with standard
deviations of $5^\circ$ and $15^\circ$ respectively).  Distribution B
can be fit by two Gaussians of means $-63^\circ$ and $-82^\circ$ (with
standard deviations of $7^\circ$ and $9^\circ$ respectively).  The
vectors are on average shifted by $-30^\circ$ by the improved data
reduction techniques of POLPACK.  The effect is restricted to the
MMS-9 region.
\label{shift}
}

\figcaption[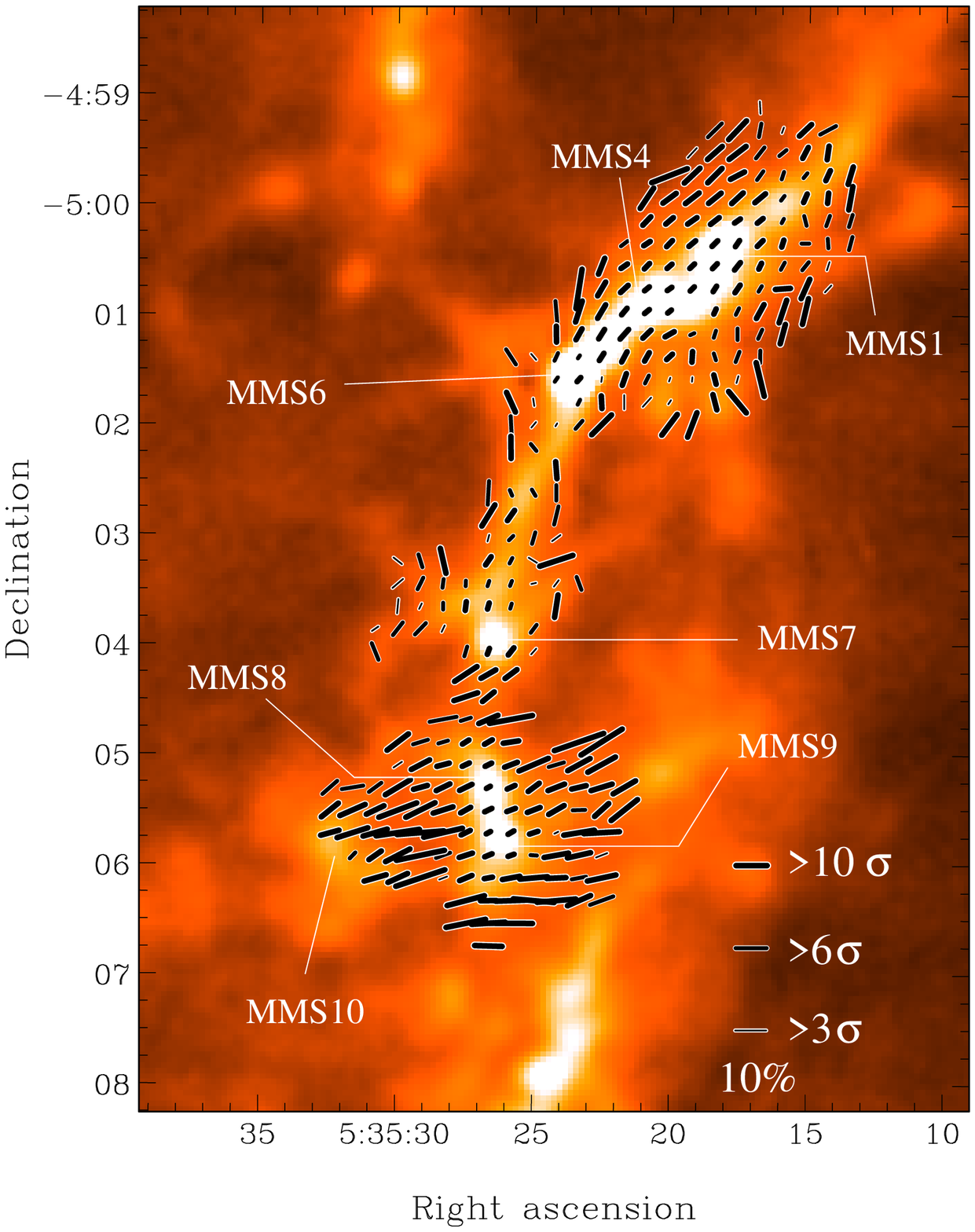]{850 \micron\ polarization pattern across OMC-3.
A portion of the 850 \micron\ intensity map of \citet{joh99} is shown
in colored greyscale.  The greyscale range is $-1.5$ to $3.5\sigma$.
The polarization mapping covers only a portion of the area shown.
Polarization data were sampled at 3\arcsec\ and have been binned to
12\arcsec\ (slightly less than the JCMT 850 \micron\ beamwidth of
14\arcsec).  The polarization vectors plotted all have percentage
polarization $> 1$\%, an uncertainty in polarization percentage $<
1$\% and a total intensity three times that of the sky bolometer level
and $\sim 10$\% of the faintest peak, MMS7.  The thinnest vectors have
a signal-to-noise in polarization percentage, $\sigma_p > 3$, while
the medium thickness vectors have $\sigma_p > 6$.  Most vectors are
bold and have $\sigma_p > 10$.  These vectors are accurate in position
angle to better than $10^\circ$, $4.8^\circ$ and $2.9^\circ$,
respectively.  The central region of MMS7 and the region south of MMS6
are devoid of vectors since the polarization percentages there are
less than 1\%.  The mean polarization percentage of the plotted
vectors is 5.0\% in 286 vectors.  The coordinates of the map are J2000.
\label{mapnew}
}

\figcaption[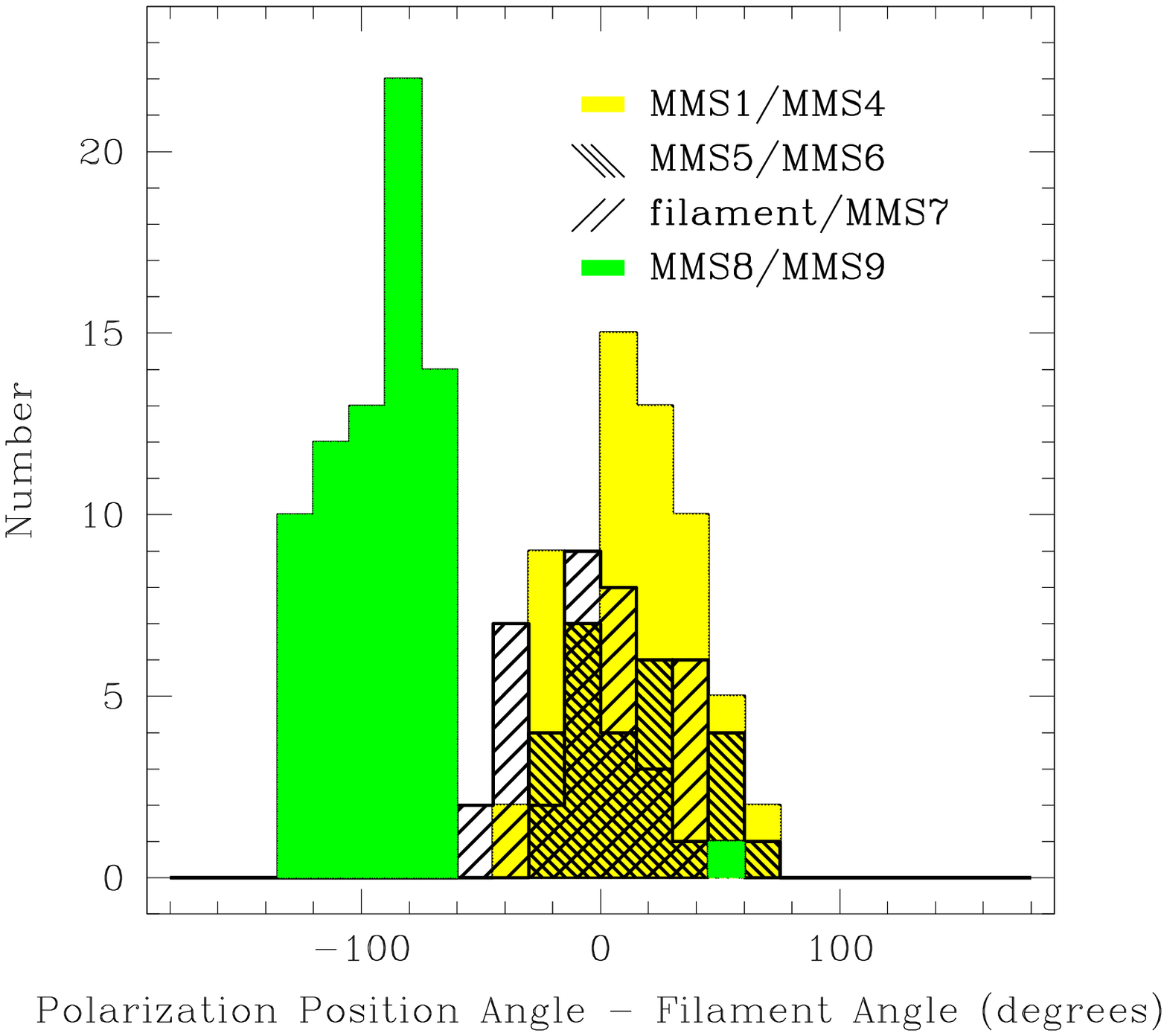]{The relative angles between the filament
orientation and the position angles of polarization vectors (with
$\sigma_p > 6$) along radial cuts to the filament are shown for four
regions of the OMC-3 map.  Only vectors with $\sigma_p > 6$ are used.
The filament orientation is derived from a cubic spline fit to the
intensity of OMC-3.  For three of the four subregions, the offsets are
centered around zero.  For the region around MMS8 and MMS9, a Gaussian
fit to this profile yields a mean and standard deviation of $-94^\circ
\pm 19^\circ$.
\label{offsetpa}
}

\figcaption[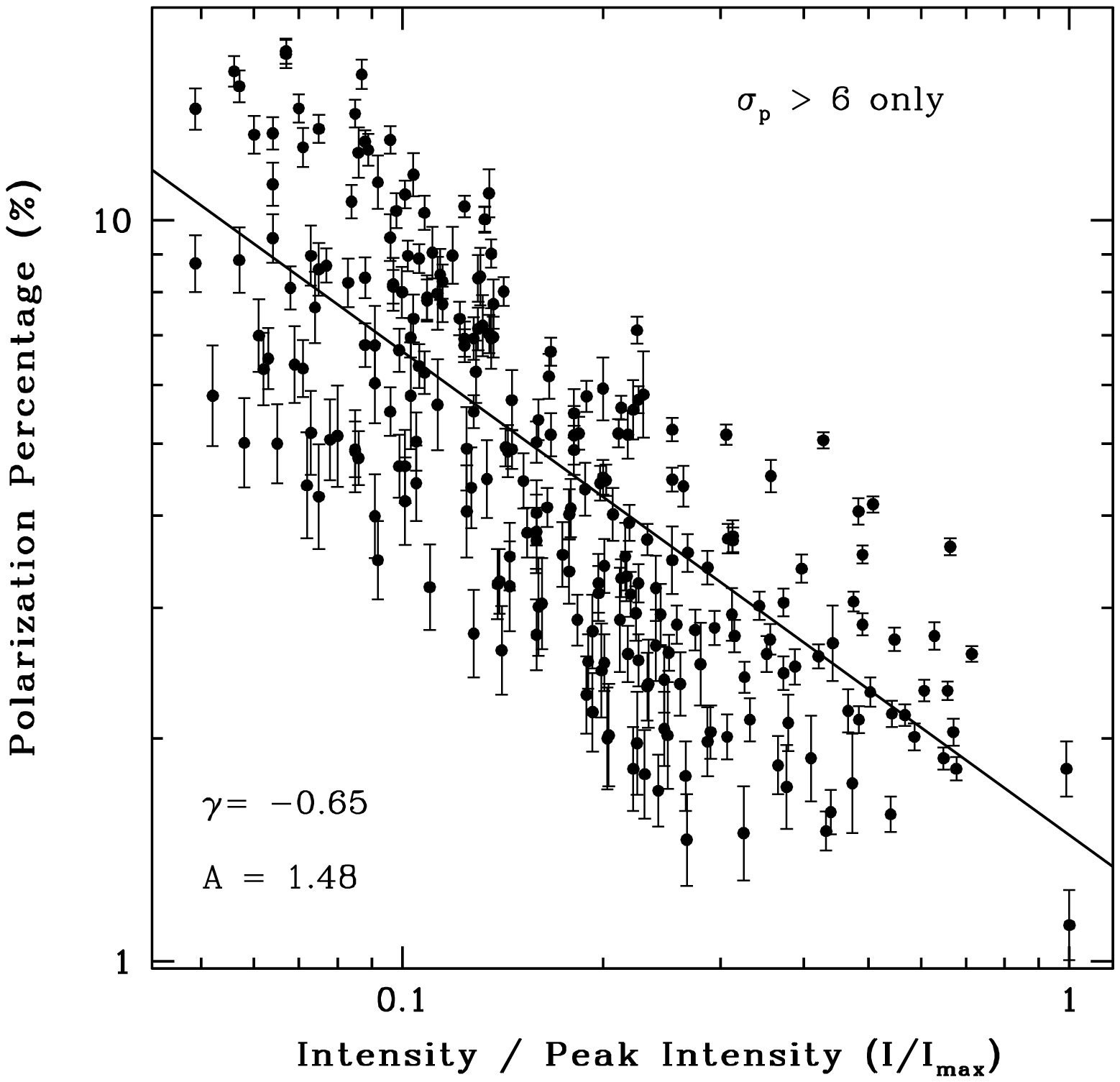]{A logarithmic plot of polarization percentage
versus intensity (scaled by the maximum intensity point) reveals that
higher intensities have systematically lower polarizations.  Values
plotted are those on Figure \ref{mapnew} which have $\sigma_p > 6$.
The decreasing trend cannot be accounted for by the uncertainties
shown here.  A $\chi^2$ power law fit of the form $p= A \times
I^\gamma$ yields the $A$ and $\gamma$ parameters recorded on the
log-log plot.  A slope of $-0.65$ effectively characterizes these
data.
\label{pvsI}
}

\figcaption[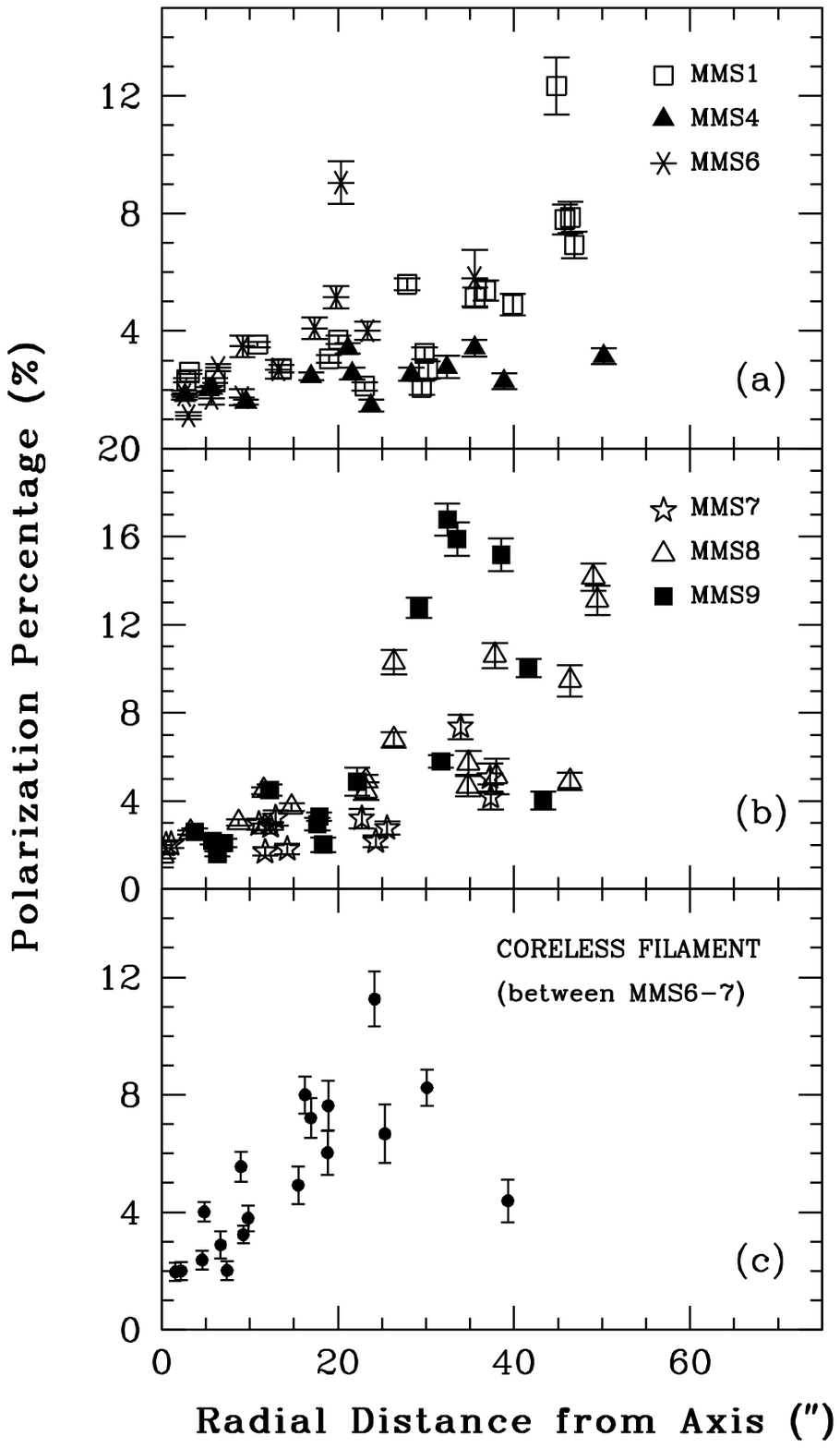]{Plots showing the polarization
percentages as a function of radial distance from the filament axis
toward different regions of OMC-3.  Depolarization is exhibited toward
the filament axis in all cases.  Panels (a) and (b) show data around
all six distinguishable cores. Panel (c) shows the coreless region
between MMS6 and MMS7.  The depolarization effect is particularly
strong for MMS8 and MMS9.  Only vectors with $\sigma_p>6$ are used.
\label{pvsr}
}

\figcaption[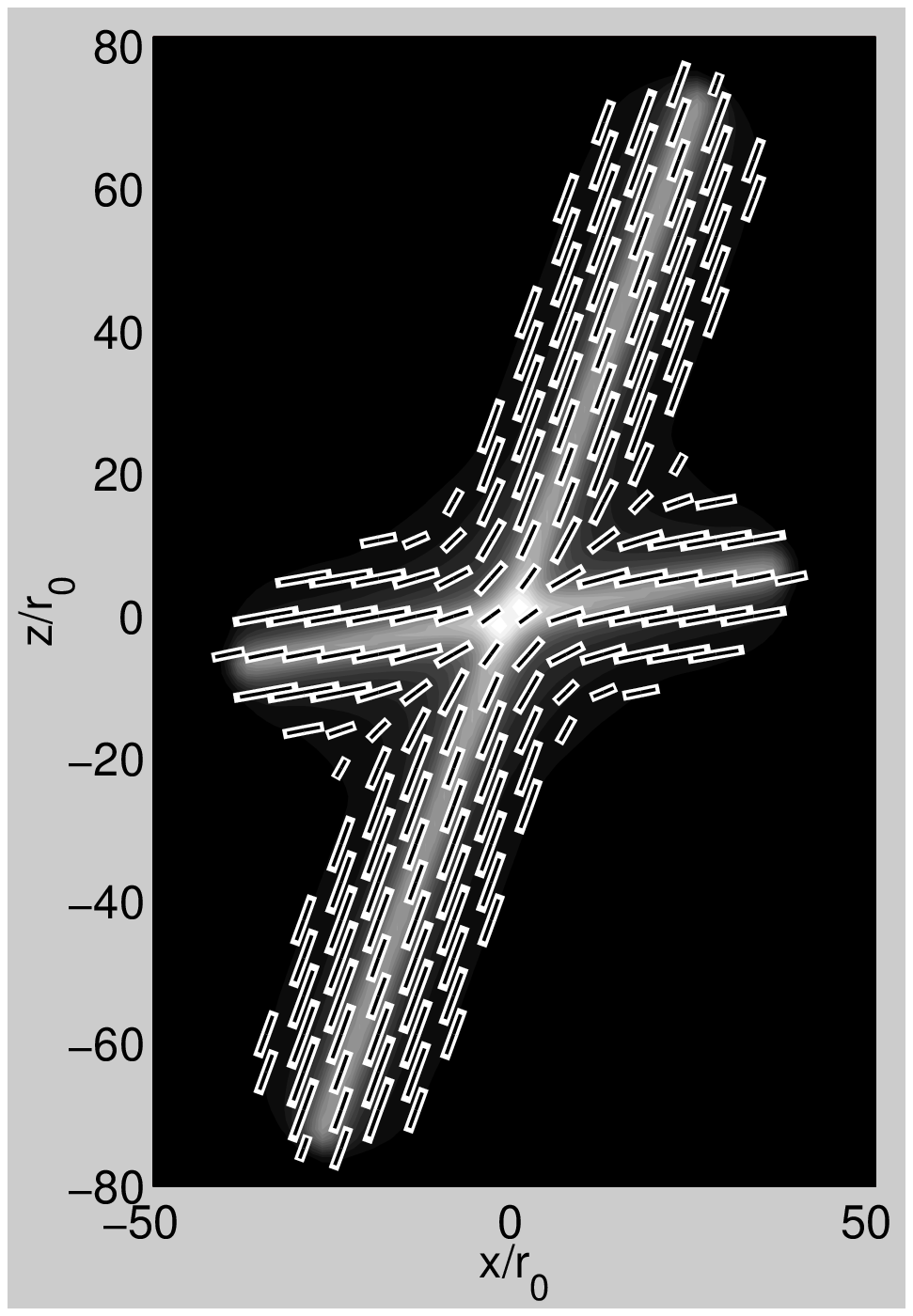]{A qualitative model of the
polarization pattern produced by a crossing of two filaments threaded
by helical magnetic fields.  The second (roughly east-west) filament
has only half the central density of the main filament, which is meant
to represent the \intfil.  Both filaments lie in the plane of the sky.
The model has been convolved with a Gaussian where the beamsize is one
fifth the filament diameter.  Vectors shown have $p > 0.05 p_{max}$
and $I> 0.05 I_{max}$ where $p_{max}=10$\%.
\label{crossed}
}

\figcaption[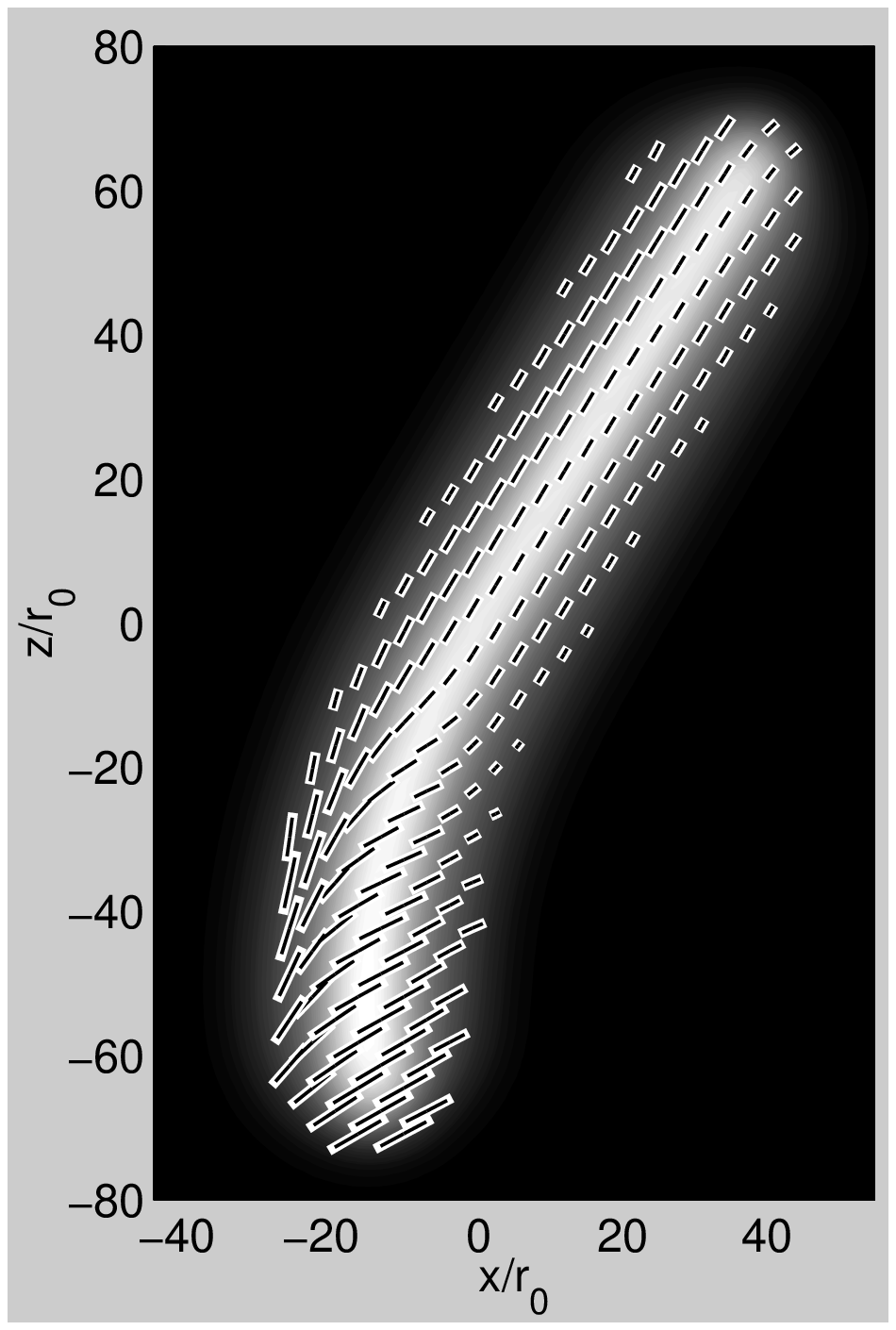]{A qualitative model of the polarization pattern
produced by a magnetized filamentary cloud where the southern half of
the filament has been bent into an arc with a radius of curvature of
6.4 times the filament's diameter.  The whole filament is inclined to
the plane of the sky at an angle of $20^\circ$ and then rotated in the
plane of the sky by $225^\circ$.  The model is convolved with the same
Gaussian beam as Figure \ref{crossed}, and the constraints on the
vectors displayed are identical.  The bending of the filament breaks
the symmetry of the models presented in \citet{fp00c} causing the
polarization pattern from the inner region of the bend to dominate, as
discussed in the text.
\label{bent}
}

\figcaption[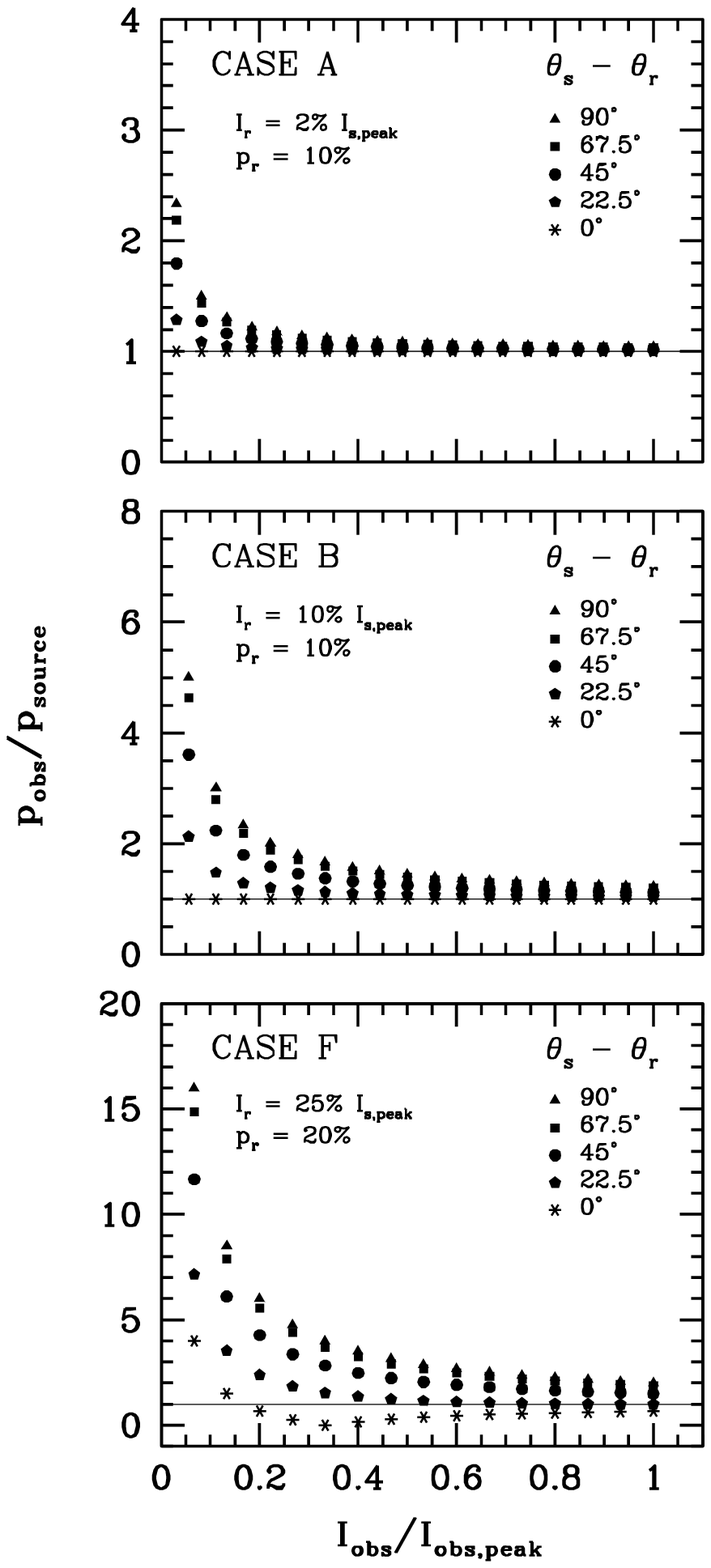]{Ratio of observed to source
polarization percentage versus observed intensity as a function of the
observed peak intensity for three of the cases identified in Table
\ref{tableA1}.  In each case, the source is polarized at a level of
10\%.  The ratios of input parameters of polarization percentage and
intensity at the reference position compared to those at the source
position are smallest for Case A and largest for Case F.  The offset
between source and reference angles, \pas\ $-$ \pachop, are labelled
in each case.  Clearly, the systematic effects due to chopping become
more significant as the polarized flux in the reference beam grows
relative to that of the source.
\label{polchop_p}
}

\figcaption[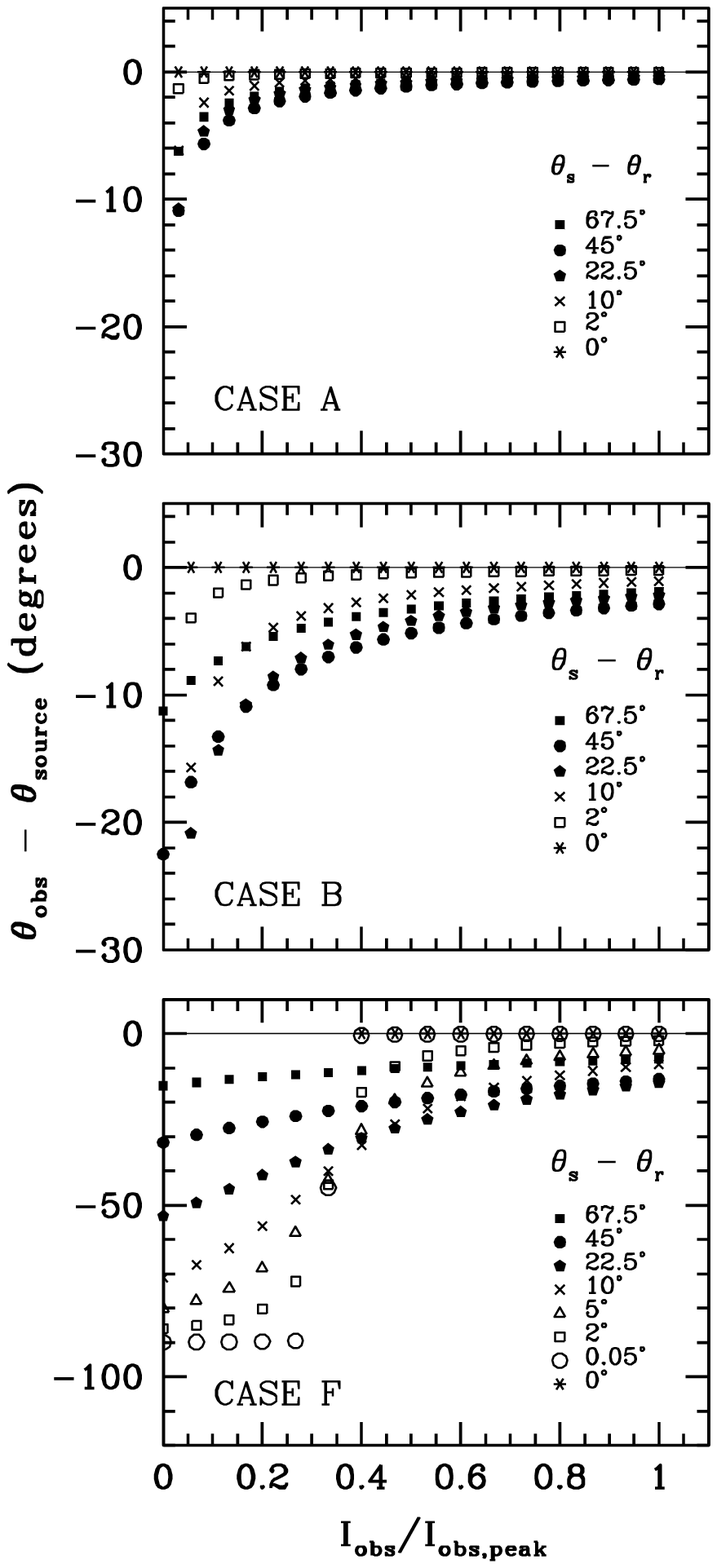]{Observed position angles at various
observed intensities (as a function of the observed flux peak) for
Cases A, B and F as described in Table \ref{tableA1}.  The assigned
position angle of the source was 0$^\circ$.  Solutions for positive
and negative offsets between source and reference position angles are
symmetric about the source position angle; hence, we show only
solutions for positive offsets here.  The solution for an offset of
90$^\circ$ is discussed in the text and is identical to the solution
for 0$^\circ$.  The offset angles, \pas\ $-$ \pachop, are labelled in
each case.  As for percentage polarization in Figure \ref{polchop_p},
the lower the ratio of total flux in the reference position to the
source, the less impact chopping has on the observed values at the
source position.
\label{polchop_pa}
}

\figcaption[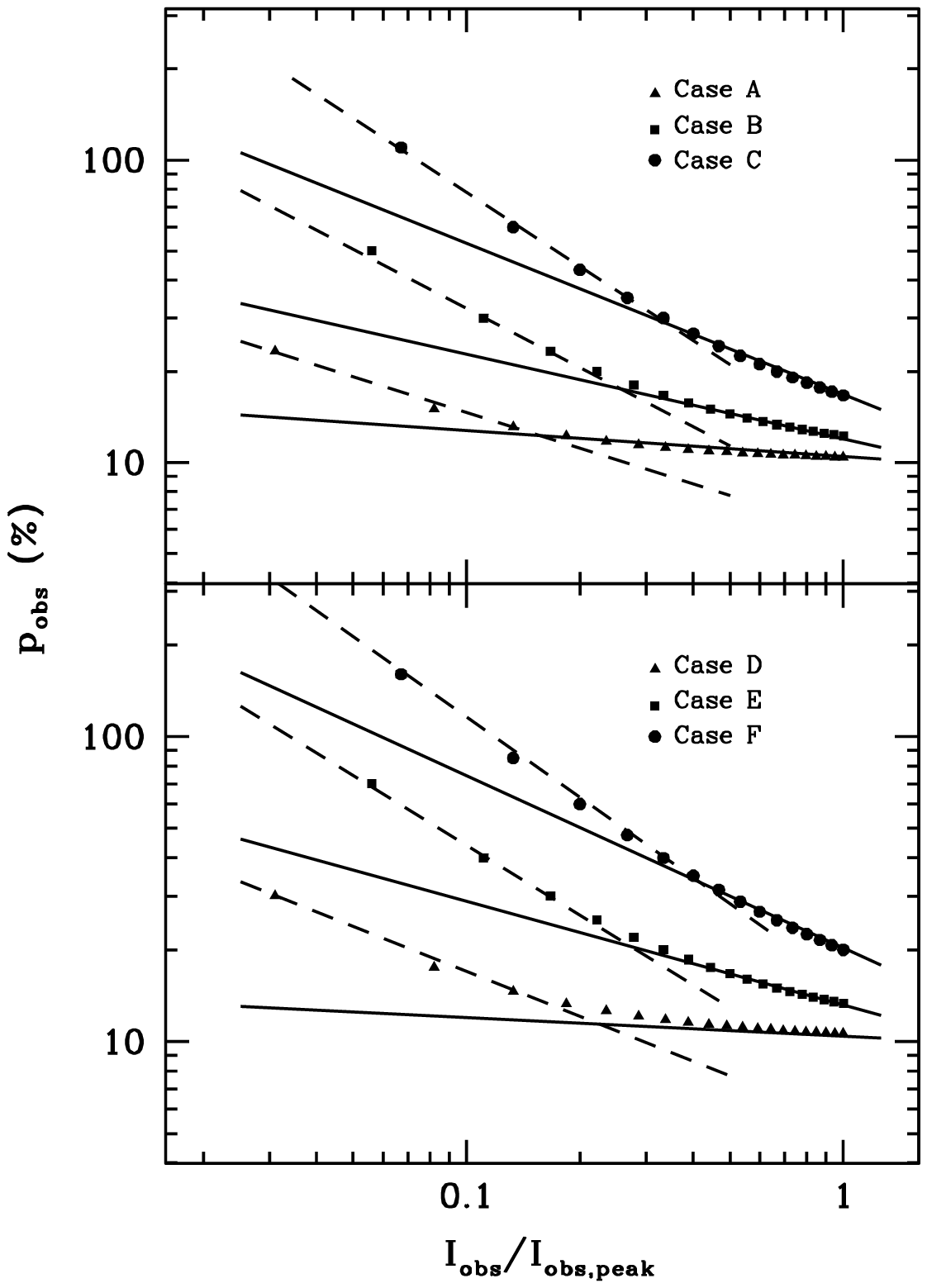]{Depolarization effects expected for the
cases described in Table \ref{tableA1}.  As Figure \ref{pvsI} shows,
decreased polarization percentage with increased intensity is a global
feature in OMC-3.  Such an effect can clearly be produced by chopping
onto a reference position with significant polarization, although the
magnitude of the slope produced diminishes as the flux of the
reference position (with respect to the source) decreases.
\label{logpI_expected}
}

\figcaption[fig11.ps]{Six data sets of the MMS8-MMS9 region
are shown, each reduced with (right) and without (left) sky
subtraction.  These maps reveal that a high degree of uniformity in
position angle can be observed before sky subtraction is performed.
These patterns can be created by variations in sky conditions during
the 12 minute polarization cycle and must be removed to reveal the
true polarization vectors of the source.
\label{appendixB_fig2}
}

\figcaption[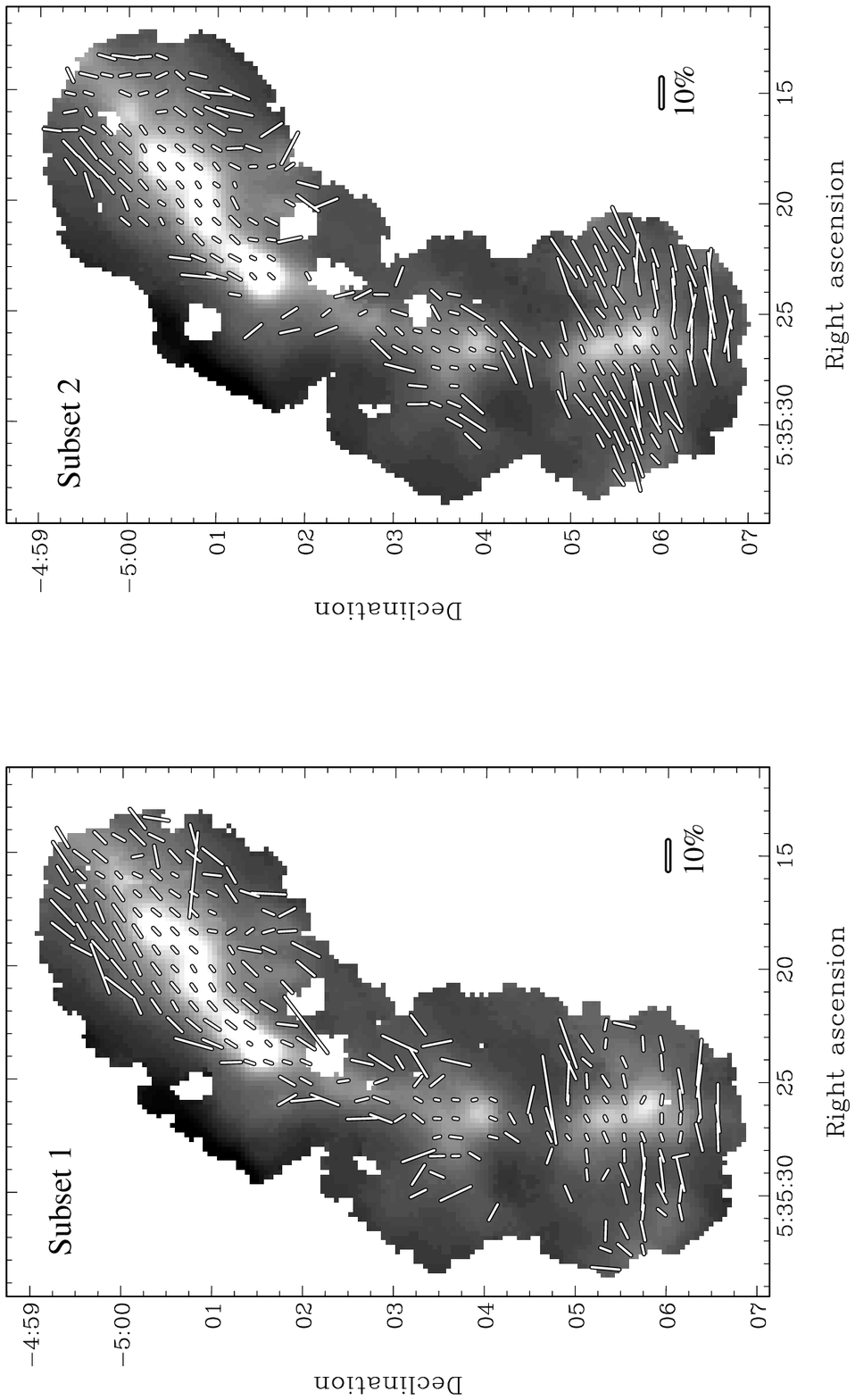]{These two figures illustrate the
consistency between two halves of the OMC-3 data set.  Polarization
vectors plotted have $p>1$\%, an uncertainty in polarization
percentage, $dp$, $<1.4$\% and $p/dp>4.2$.
\label{appendixC_fig}
}

\newpage
\plotone{fig1.ps}

\newpage
\plotone{fig2.eps}

\newpage
\plotone{fig3.ps}

\newpage
\plotone{fig4.ps}

\newpage
\plotone{fig5.ps}

\newpage
\begin{figure}[htb!]
\vspace*{8cm}
\includegraphics{fig6.eps}
\end{figure}

\clearpage
\begin{figure}[htb!]
\vspace*{8cm}
\includegraphics{fig7.eps}
\end{figure}

\newpage
\plotone{fig8.ps}

\newpage
\plotone{fig9.ps}
\newpage
\plotone{fig10.ps}

\newpage
\plotone{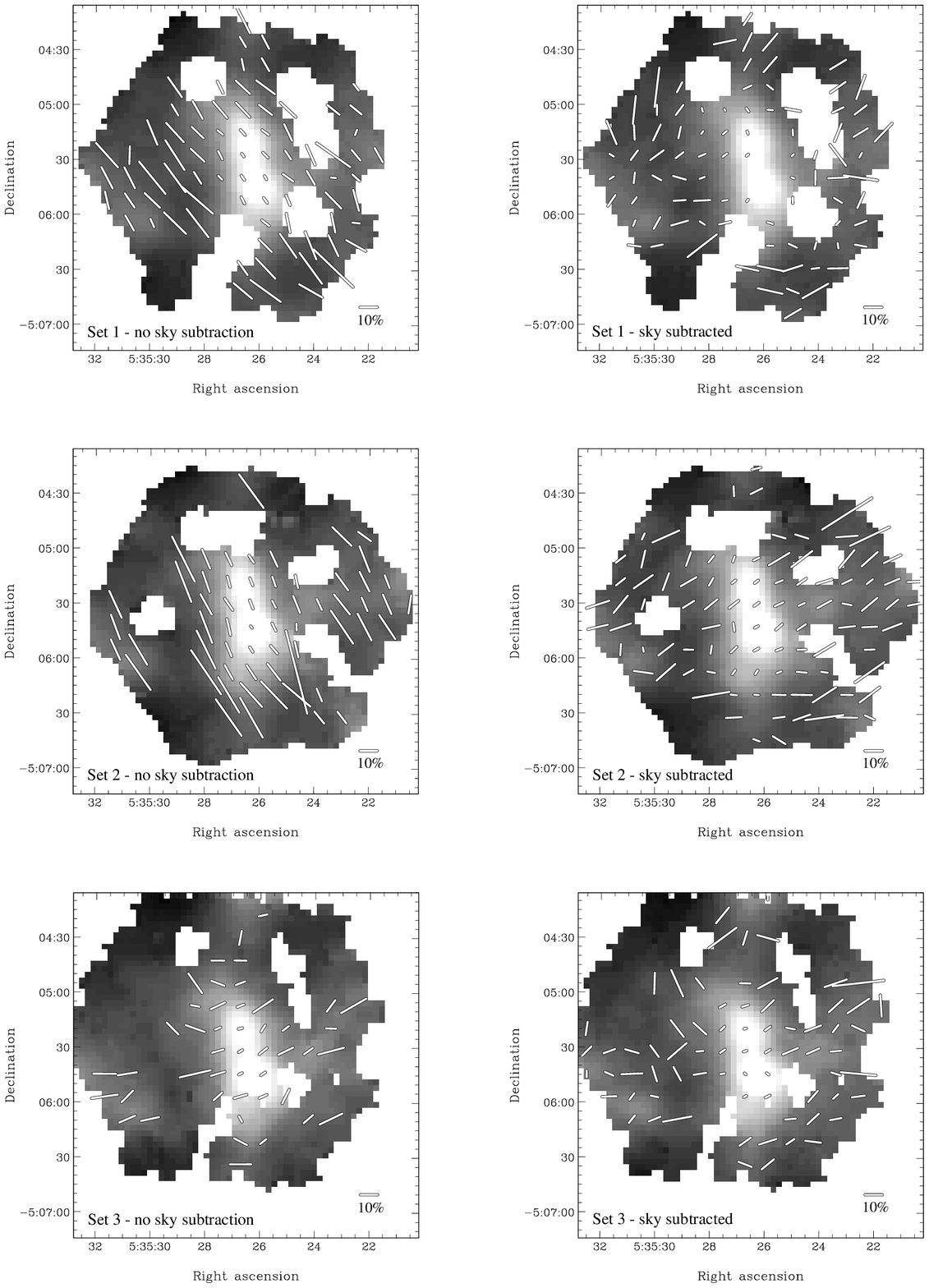}

\newpage
\plotone{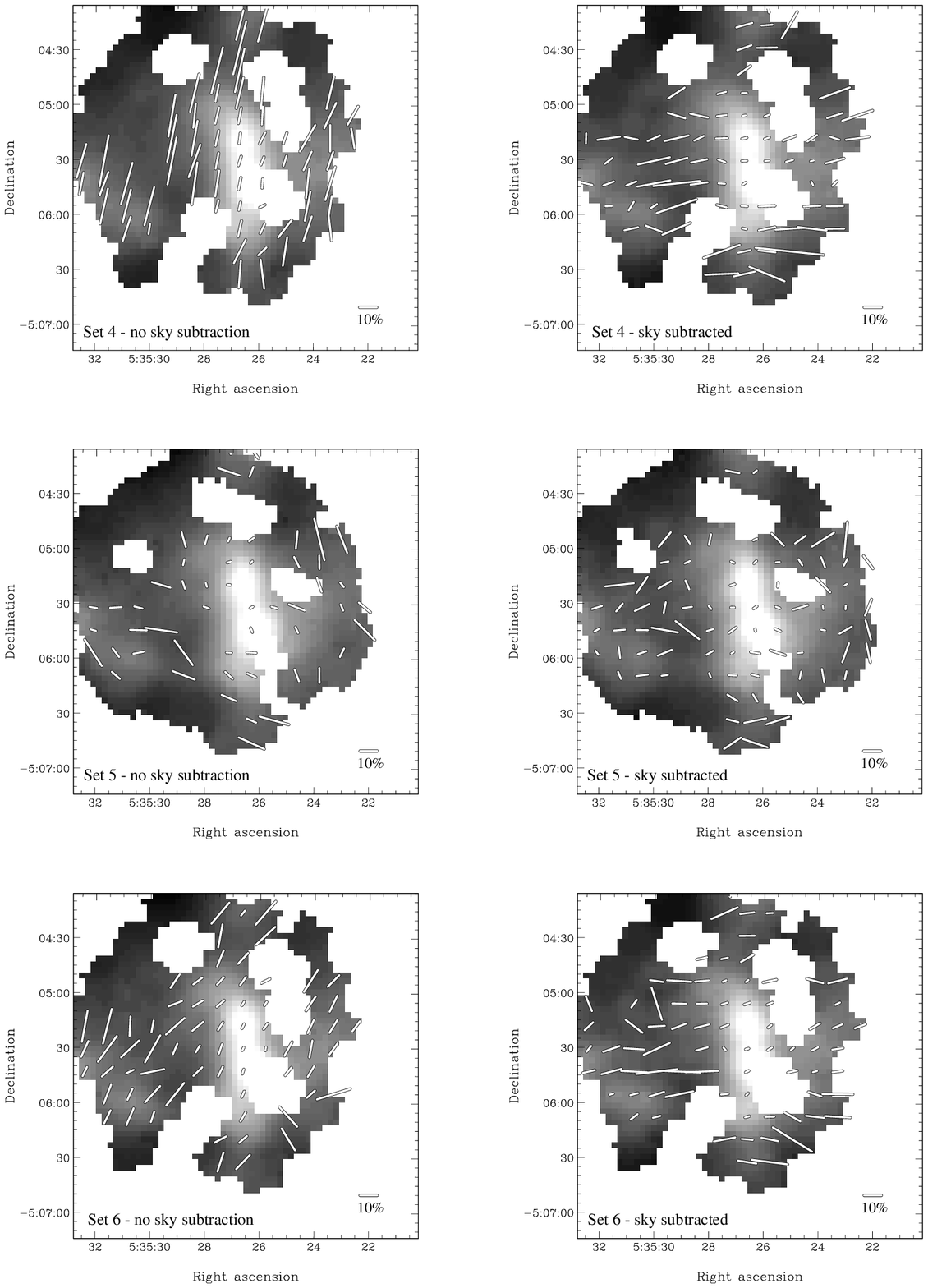}

\newpage
\plotone{fig12.ps}

\end{document}